\documentclass[aps,prb,reprint,superscriptaddress]{revtex4-1}
\usepackage{amsmath,amsthm,amssymb,graphicx,bm,setspace,calc,float,xr,natbib,epstopdf,hyperref,wasysym,cleveref}
\usepackage[svgnames]{xcolor}
\hypersetup{hidelinks,colorlinks=true,allcolors=DarkBlue,pdfpagemode=UseNone}

\begin{document}
\newcommand{\figSize}{0.5}
\newcommand{\GRad}{\Gamma_{\mathrm{Rad}}}
\newcommand{\GMix}{\Gamma_{\mathrm{Mix}}}
\newcommand{\GTTwo}{\Gamma_{T_2}}
\newcommand{\GAdd}{\Gamma_\mathrm{Add}}
\newcommand{\GISC}{\Gamma_\mathrm{ISC}}
\newcommand{\tAdd}{\tau_\mathrm{Add}}
\newcommand{\tRad}{\tau_\mathrm{Rad}}
\newcommand{\tRabi}{\tau_\mathrm{Rabi}}
\newcommand{\GAone}{\Gamma_{A_1}}
\newcommand{\GAtwo}{\Gamma_{A_2}}
\newcommand{\GEonetwo}{\Gamma_{E_{1,2}}}
\newcommand{\As}{|^1A_1\rangle}
\newcommand{\At}{\,^3A_2}
\newcommand{\Es}{|^1E_{1,2}\rangle}
\newcommand{\Et}{\,^3E}
\newcommand{\Eonetwo}{|E_{1,2}\rangle}
\newcommand{\Eone}{|E_1\rangle}
\newcommand{\Etwo}{|E_2\rangle}
\newcommand{\Ex}{|E_x\rangle}
\newcommand{\Ey}{|E_y\rangle}
\newcommand{\Exy}{|E_{x,y}\rangle}
\newcommand{\Aone}{|A_1\rangle}
\newcommand{\Atwo}{|A_2\rangle}
\newcommand{\Aonetwo}{|A_{1,2}\rangle}
\newcommand{\dAoa}{\delta_{A_1^a}}
\newcommand{\dAob}{\delta_{A_1^b}}
\newcommand{\dEoa}{\delta_{E_1^a}}
\newcommand{\dEob}{\delta_{E_1^b}}
\newcommand{\dEta}{\delta_{E_2^a}}
\newcommand{\dEtb}{\delta_{E_2^b}}
\newcommand{\dEota}{\delta_{E_{1,2}^a}}
\newcommand{\pAoa}{\hat{P}_{A_1^a}}
\newcommand{\pAob}{\hat{P}_{A_1^b}}
\newcommand{\pEoa}{\hat{P}_{E_1^a}}
\newcommand{\pEob}{\hat{P}_{E_1^b}}
\newcommand{\pEta}{\hat{P}_{E_2^a}}
\newcommand{\pEtb}{\hat{P}_{E_2^b}}
\newcommand{\msz}{m_s=0}
\newcommand{\mso}{|m_s|=1}
\newcommand{\wA}{\omega_{A_1}}
\newcommand{\wE}{\omega_E}

\newcommand{\GRadx}{\Gamma_\mathrm{Rad}^{(x)}}
\newcommand{\GRady}{\Gamma_\mathrm{Rad}^{(y)}}
\newcommand{\Gxy}{\Gamma_\mathrm{Mix}^{(x)}}
\newcommand{\Gyx}{\Gamma_\mathrm{Mix}^{(y)}}
\newcommand{\GISCx}{\Gamma_{\mathrm{ISC},E_x}}
\newcommand{\GISCone}{\Gamma_{\mathrm{ISC},\pm1}}
\newcommand{\GISCzero}{\Gamma_{\mathrm{ISC},0}}

\title{State-selective intersystem crossing in nitrogen-vacancy centers}

\author{M. L. Goldman}
\email[]{mgoldman@physics.harvard.edu}
\affiliation{Department of Physics, Harvard University, Cambridge, Massachusetts 02138, USA}

\author{M. W. Doherty}
\affiliation{Laser Physics Centre, Research School of Physics and Engineering, Australian National University, Australian Capital Territory 0200, Australia}

\author{A. Sipahigil}
\affiliation{Department of Physics, Harvard University, Cambridge, Massachusetts 02138, USA}

\author{N. Y. Yao}
\affiliation{Department of Physics, Harvard University, Cambridge, Massachusetts 02138, USA}

\author{S. D. Bennett}
\affiliation{Department of Physics, Harvard University, Cambridge, Massachusetts 02138, USA}

%

\author{N. B. Manson}
\affiliation{Laser Physics Centre, Research School of Physics and Engineering, Australian National University, Australian Capital Territory 0200, Australia}

\author{A. Kubanek}
\affiliation{Department of Physics, Harvard University, Cambridge, Massachusetts 02138, USA}

\author{M. D. Lukin}
\affiliation{Department of Physics, Harvard University, Cambridge, Massachusetts 02138, USA}

\begin{abstract}
The intersystem crossing (ISC) is an important process in many solid-state atomlike impurities.  For example, it allows the electronic spin state of the nitrogen-vacancy (NV) center in diamond to be initialized and read out using optical fields at ambient temperatures. This capability has enabled a wide array of applications in metrology and quantum information science.  Here, we develop a microscopic model of the state-selective ISC from the optical excited state manifold of the NV center.  By correlating the electron-phonon interactions that mediate the ISC with those that induce population dynamics within the NV center's excited state manifold and those that produce the phonon sidebands of its optical transitions, we quantitatively demonstrate that our model is consistent with recent ISC measurements.  Furthermore, our model constrains the unknown energy spacings between the center's spin-singlet and spin-triplet levels. Finally, we discuss prospects to engineer the ISC in order to improve the spin initialization and readout fidelities of NV centers.
\end{abstract}
\maketitle


\section{Introduction}
\label{sec.Introduction}

The nitrogen-vacancy (NV) center in diamond has recently been applied to a diverse range of room-temperature applications in metrology and quantum information science.  NV centers have been used to sense such quantities as the temperature and magnetic fields of living cells \cite{Kucsko2013,LeSage2013}, the magnetic field of a single electron spin \cite{Grinolds2013}, the magnetic field noise of a few molecules \cite{Ermakova2013,Sushkov2013}, an electric field equivalent to a single electron at a distance of 35 nm \cite{Dolde2011}, and pressures of up to 60 GPa \cite{Doherty2014}.  NV centers can also function as room-temperature quantum registers, where the electronic spin can be coherently coupled to the nitrogen nucleus \cite{Dutt2007} by decoherence-protected gates \cite{VanderSar2012} and proximal $^{13}\mathrm{C}$ nuclear spins with second-long coherence times \cite{Maurer2012a} can serve as storage qubits.  These applications depend on the optical initialization and readout of the NV center's electronic spin, techniques which are enabled by the intersystem crossing (ISC) mechanism.

The ISC mechanism, which refers to nonradiative transitions between states of different spin multiplicity, has been previously investigated both theoretically \cite{Manson2006} and by indirect experimental methods, such as measurements of spin-resolved fluorescence lifetimes \cite{Toyli2012,Batalov2008,Robledo2011} and spin dynamics under nonresonant optical excitation \cite{Manson2006,Robledo2011,Tetienne2012}. While the previous theoretical investigation of the ISC mechanism established that it involves both spin-orbit and electron-phonon interactions, it did not provide a detailed microscopic model of the mechanism. Recently, the ISC rates from each of the fine structure states of the center's optical excited level have been measured at cryogenic temperatures \cite{Goldman2014}. These new measurements complete a comprehensive experimental picture of the ISC from the optical excited level, thereby motivating new theoretical efforts to develop a detailed model of the ISC mechanism.

In this paper, we present such a microscopic model of the state-selective ISC from the optical excited state manifold of the NV center.  We show that the electron-phonon interactions that mediate the ISC are closely linked with those that induce population dynamics within the NV center's excited state manifold and those that produce the phonon sidebands (PSBs) of its optical transitions.  This correspondence enables us to use recent measurements of the phonon-induced population mixing rate \cite{Goldman2014} and the PSB of the visible transition \cite{Kehayias2013} as experimental inputs to our model, and we quantitatively demonstrate that our model is consistent with recent ISC measurements.  Additionally, our model constrains the unknown energy spacings between the center's spin-singlet and spin-triplet levels to spectral regions that may be probed in future measurements. The identification of these energy spacings will resolve the most significant unknown aspect of the center's electronic structure. Finally, we discuss how our new understanding yields prospects to engineer the ISC to improve the spin initialization and readout fidelities of NV centers at room temperature.

This work is structured as follows: In Sec. \ref{sec.ISC Model}, we describe the NV center, present the formalism of our model of the ISC mechanism, and explicitly calculate the ISC rates from different states in the $\Et$ manifold.  We also calculate the phonon-induced mixing rate between $\Ex$ and $\Ey$, which enables us to extract the NV-phonon coupling strength from prior experimental observations.  In Sec. \ref{sec.Comparison to Measurements}, we compare the results of our model to recently measured state-selective ISC rates at cryogenic temperatures, which enables us to place bounds on the energy spacing between the spin-triplet and -singlet states.  In Sec. \ref{sec.Extension of Model to High Temperatures}, we extend our model to higher temperature and show that it is consistent with previous observations of spin-dependent fluorescence lifetimes.  We conclude in Sec. \ref{sec.Conclusion} by suggesting future theoretical and experimental directions.

\section{ISC Model}
\label{sec.ISC Model}

\subsection{Level Structure of NV Center}

\begin{figure}[h]
\begin{center}
\includegraphics[width=\columnwidth]{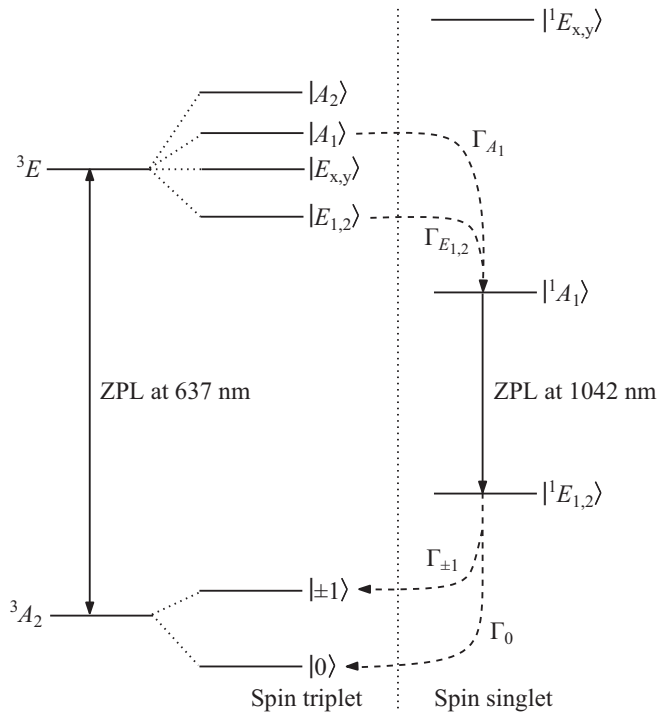}
\caption{Electronic structure of the NV center, including the fine structure states of the spin-triplet levels, the ZPL transitions of the center's visible and infrared resonances (solid arrows), and the state-selective nonradiative ISCs between the spin-triplet and -singlet levels (dashed arrows). $\Gamma_X$ denote the rates of the individual ISC transitions.}
\label{fig.NV Electronic States}
\end{center}
\end{figure}

The NV center is a point defect of $C_{3v}$ symmetry that consists of a substitutional nitrogen atom adjacent to a vacancy in the diamond lattice. In the negative charge state, six electrons occupy the four dangling $sp^3$ orbitals of the nitrogen and the vacancy's three nearest-neighbor carbon atoms \cite{Maze2011,Doherty2011}. The four $sp^3$ atomic orbitals linearly combine to form four symmetry-adapted molecular orbitals ($a_1^\prime$, $a_1$, $e_x$, $e_y$), such that the ground electronic configuration is $a_1^{\prime2}a_1^2e^2$.  The ground electronic state (labeled $\At$) is an orbital-singlet, spin-triplet manifold (see Fig. \ref{fig.NV Electronic States} for the NV center's electronic structure).  Within $\At$, there is a doublet of states with $m_s=\pm1$ spin projection along the N-V axis (labeled $|\pm1\rangle$) located 2.87 GHz above one state with $\msz$ (labeled $|0\rangle$). The ground electronic configuration also contains an orbital-singlet, spin-singlet state $\As$ that is higher in energy than the orbital-doublet, spin-singlet states $\Es$. The $\As$ and $\Es$ singlet states are coupled by an optical transition with a zero-phonon line (ZPL) at 1042 nm \cite{Rogers2008}. The first excited electronic configuration ($a_1^{\prime2}a_1e^3$) consists of the optical excited orbital-doublet, spin-triplet manifold $\Et$, as well as the orbital doublet, spin-singlet states $|^1E_{x,y}\rangle$. The $\Et$ manifold is coupled to the $\At$ ground state manifold by an optical transition with a ZPL at 637 nm.  The two orbital states and three spin states of the $\Et$ manifold combine to give a total of six fine structure states: two (labeled $\Ex$ and $\Ey$) have zero spin angular momentum projections, while the other four (labeled $|A_1\rangle$, $|A_2\rangle$, $|E_1\rangle$, and $|E_2\rangle$) are entangled states of nonzero spin and orbital angular momentum projections.

The spin dynamics of the NV center under optical illumination are driven by radiative transitions between states of the same spin multiplicity as well as nonradiative ISCs between states of different spin multiplicity (see Fig. \ref{fig.NV Electronic States}). There are two distinct ISCs: from the optical excited $\Et$ manifold to the higher energy singlet state $\As$, and from the lower energy singlet states $\Es$ to the ground $\At$ manifold. The ISC from the $\Et$ manifold may occur from either $\Aone$ with rate $\GAone$ or from $\Eonetwo$ with rate $\GEonetwo\approx\GAone/2$ \cite{Goldman2014}. While direct ISCs from $\Exy$ and $\Atwo$ are not forbidden, their rates have been established to be negligible compared to $\GAone$ and $\GEonetwo$ at cryogenic temperatures \cite{Goldman2014}.  This hierarchy reflects the directness of the physical process that couples each $\Et$ state to the singlet states: the ISCs from $\Aone$ and $\Eonetwo$ are respectively mediated by first- and second-order processes, as discussed in Secs. \ref{sec.A1 ISC} and \ref{sec.E12 ISC}, and the lowest-order allowed ISC processes from $\Exy$ and $\Atwo$ would be third-order, as discussed in Ref. \onlinecite{Goldman2014}.  The ISC from the $\Et$ manifold is thus highly state-selective.  The ISC to the $\At$ manifold occurs from $\Es$ to either $|0\rangle$ with rate $\Gamma_0$ or $|\pm1\rangle$ with rate $\Gamma_{\pm1}$. The ratio $\Gamma_0/\Gamma_{\pm1}$ appears to vary between centers with observed values in the range 1.1--2 \cite{Robledo2011,Tetienne2012}. While further investigations are required, it is clear that there is no strong state selectivity of the ISC to the $\At$ manifold.

Under optical excitation, the radiative transitions between the $\At$ and $\Et$ manifolds conserve electronic spin-projection, whereas the highly state-selective ISC transitions from the $\Et$ manifold to $\As$ preferentially depopulate $|A_1\rangle$ and $|E_{1,2}\rangle$, and the ISC transitions from $\Es$ to the $\At$ manifold serve to repopulate the ground state for the next optical cycle. The preferential nonradiative depopulation of $|A_1\rangle$ and $|E_{1,2}\rangle$ gives rise to optical spin readout because these states have lower quantum yield \cite{Doherty2013a}. Since the nonradiative repopulation of the $\At$ manifold from $\Es$ does not appear to be comparably state-selective, the preferential nonradiative depopulation of $|A_1\rangle$ and $|E_{1,2}\rangle$ is also predominately responsible for the optical initialization of the electronic spin into $|0\rangle$ \cite{Doherty2013a}. Given the central role of the ISC from the optical excited $\Et$ manifold to $\As$ in the NV optical-spin cycle, we focus our attention on developing a microscopic model of this ISC.

\subsection{ISC Mechanism}

\begin{figure}[h]
\begin{center}
\includegraphics[width=\columnwidth]{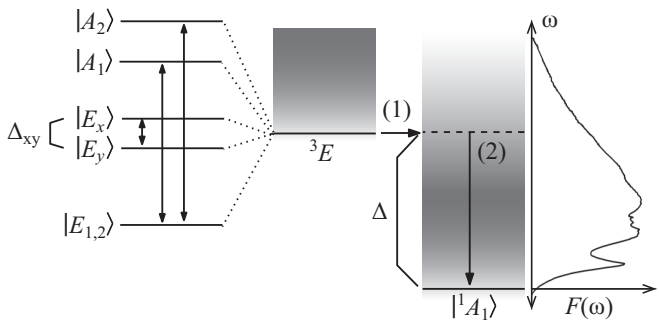}
\caption{Schematic of (left) the phonon-induced mixing within the $\Et$ manifold and (right) the stages of the ISC from the $\Et$ manifold to the $\As$ state. The phonon mixing transitions within the $\Et$ manifold are depicted by solid arrows. $\Delta_{xy}$ is the strain-induced splitting of the $|E_{x,y}\rangle$ states. The shaded regions denote the quasi-continua of the vibrational levels of $\Et$ and $\As$.  The stages of the ISC, which are described in the text, are denoted by the arrows labeled (1) and (2).  The ISC vibrational overlap function $F\hspace{-1.5 pt}\left(\omega\right)$, as approximated by the visible emission PSB, is depicted on the far right. This provides a visual representation of how the $\Et$--$\As$ energy separation $\Delta$ influences the ISC rate.}
\label{fig.ISCdiagram}
\end{center}
\end{figure}

ISCs from the $\Et$ manifold to the $\As$ state occur in two stages (see Fig. \ref{fig.ISCdiagram}): (1) an energy-conserving transition from the initial state within the $\Et$ manifold to a resonant excited vibrational level of the $\As$ state, and (2) relaxation of the excited vibrational level to the ground (or thermally occupied) vibrational level of $\As$. The first stage requires a change in both the electron spin and orbital states as well as the lattice vibrational state, and is thus mediated by a combination of spin-orbit (SO) and electron-phonon interactions.  The second stage is mediated by phonon-phonon interactions, which enables the vibrational excitation to dissipate into propagating phonon modes. As the vibrational relaxation occurs on picosecond timescales \cite{Huxter2013}, the ISC rate is defined by the initial electronic transition.

The model of these transitions requires the description of the quasi-continuum of vibronic levels of the $\Et$ and $\As$ electronic states.  Coulombic interactions between the NV center's electrons and proximal nuclei induce a local $A_1$-symmetric deformation of the diamond lattice.  Because this Coulomb force depends on the electronic charge density, the equilibrium lattice configuration depends on the NV center's electronic orbital state \cite{Gali2011,Zhang2011}.  Adopting the Born-Oppenheimer and harmonic approximations, the dependence of the NV center's energy on the positions of proximal nuclei can be modeled as a series of quadratic potentials centered on a common electronic state-dependent equilibrium point.  Thus, the vibronic levels of each electronic state take the direct product form: $\{|E_{1,2}\rangle, |E_{x,y}\rangle, |A_1\rangle, |A_2\rangle\}\otimes\{|\chi_{\nu_n}\rangle\}$ and $\As\otimes\{|\chi_{\nu_n}^\prime\rangle\}$, where $|\chi_{\nu_n}\rangle$ and $|\chi_{\nu_n}^\prime\rangle$ are the $n^\mathrm{th}$ vibrational levels of $\Et$ and $\As$, respectively, with vibrational energies $\nu_n$ \cite{Bransden2003,Stoneham2001}.  Here, $n$ denotes the set of occupation numbers $\{m_i\}$ of all vibrational modes of the specified electronic state.  The $n^\mathrm{th}$ vibrational level and energy are then given by $|\chi_{\nu_n}\rangle = \prod_i |m_i\rangle$ and $\nu_n = \sum_i m_i \, \omega_i$, respectively, where $\omega_i$ is the energy of the $i^\mathrm{th}$ vibrational mode.

Electron-phonon interactions with $A_1$-symmetric phonon modes do not couple electronic states, but allow transitions between the vibrational levels of each electronic state. As a consequence, there exist non-zero overlaps $|\langle \chi_{\nu_n} | \chi^\prime_{\nu_{n^\prime}} \rangle|^2$ of different vibrational levels of $\Et$ and $\As$.  Since the $\As$ state belongs to the same electronic configuration as the ground $\At$ manifold, their electron densities are similar, and thus the vibrational overlaps $|\langle \chi_{\nu_n} | \chi^\prime_{\nu_{n^\prime}} \rangle|^2$ are expected to be well approximated by those between the $\Et$ and $\At$ manifolds that are observed in the visible emission PSB \cite{Kehayias2013}.

Electron-phonon interactions with $E$-symmetric phonon modes couple electronic orbital states. Consequently, they can drive spin-conserving transitions between the states of the $\Et$ manifold (see Fig. \ref{fig.ISCdiagram}). The interactions with $E$-symmetric phonons within the $\Et$ manifold are described by \cite{Maze2011}
\begin{equation}
\label{eq.Electron-phonon H}
\hat{H}_\mathrm{e-p} = \sum_{p,k} \hat{V}_\mathrm{e-p}^p \, \lambda_{p,k} \left(\hat{a}_{p,k}^\dagger+\hat{a}_{p,k}\right),
\end{equation}
where
\begin{align}
\hat{V}_\mathrm{e-p}^1 = \, & |E_x\rangle\langle E_x|-|E_y\rangle\langle E_y| + \left(|E_1\rangle\langle A_1|\right. \nonumber \\
& - \left.|E_2\rangle\langle A_2| + 2|^1E_1\rangle\langle ^1A_1| + \mathrm{h.c.}\right), \nonumber \\
\hat{V}_\mathrm{e-p}^2 = \, & |E_x\rangle\langle E_y| + i|E_2\rangle\langle A_1| - i|E_1\rangle\langle A_2| \nonumber \\
& + 2|^1E_2\rangle\langle ^1A_1| +\mathrm{h.c.},
\end{align}
$\hat{a}_{p,k}^\dagger$ and $\hat{a}_{p,k}$ are the creation and annihilation operators of an $E$-symmetric phonon with wavevector $k$ and polarization $p=\{1,2\}$ \footnote{In group theoretical terms, the polarizations \unexpanded{$p=\{1,2\}$} correspond to the first and second rows of the $E$ irreducible representation.  Geometrically, phonons of these polarizations induce strain of $E_{1,2}^a$ symmetry, as defined in Ref. \onlinecite{Maze2011}, which distorts the lattice in directions that are perpendicular to the N-V axis.}, and $\lambda_{p,k}$ is the associated phononic coupling rate.

The SO interaction has two components.  The axial SO interaction [$\propto \lambda_{||}l_z s_z$, where $s_z$ ($l_z$) is the $z$-component of the electronic spin (orbital angular momentum)] defines observable aspects of the $\Et$ fine structure but does not couple $\Et$ states with $\mso$ to spin-singlet states.  On the contrary, the transverse SO interaction [$\propto \lambda_\perp \left(l_x s_x + l_y s_y\right)$] cannot be directly observed in the $\Et$ fine structure \cite{Doherty2011,Maze2011}, but gives rise to the coupling
\begin{align}
\label{eq.SO H}
\hat{H}_\mathrm{SO} = \sqrt{2} \, \hbar \, \lambda_\perp ( & |A_1\rangle\langle ^1A_1| + |E_1\rangle\langle ^1E_1|  \nonumber \\
& + i|E_2\rangle\langle ^1E_2|) + \mathrm{h.c.}
\end{align}

In the following subsections, we explicitly calculate the ISC rates from and the phonon-induced mixing rate between different $\Et$ states.  We perform this calculation by treating $\hat{H}_\mathrm{SO}$ and $\hat{H}_\mathrm{e-p}$ as time-dependent perturbations to the vibronic states of $\Et$ and $\As$ defined by electron-phonon interactions with $A_1$-symmetric phonons. The vibrational overlap function of the visible emission PSB is used to approximate the vibrational overlaps between states in the $\Et$ manifold and $\As$. We perform the calculations in the low-temperature limit applicable to the cryogenic temperatures at which the state-selective ISC rates were recently measured \cite{Goldman2014}, where only the ground vibrational levels of the $\Et$ manifold are populated prior to the ISC transitions to $\As$.  We extend our calculations to higher temperatures in Sec \ref{sec.Extension of Model to High Temperatures}.

\subsection{ISC Rate from $\Aone$}
\label{sec.A1 ISC}

Transverse SO interaction directly couples $|A_1\rangle$ with the resonant excited vibrational states of $\As$. Consequently, this rate can be calculated by the application of first-order Fermi's golden rule
\begin{equation}
\label{eq.A1 ISC 1}
\Gamma_{A_1} = 4\pi \hbar \, \lambda_\perp^2 \sum\limits_n |\langle \chi_0 | \chi^\prime_{\nu_n} \rangle|^2 \: \delta\hspace{-1 pt}\left(\nu_n - \Delta\right),
\end{equation}
where $\Delta$ is the energy spacing between $\Et$ and $\As$ (neglecting fine structure of $\Et$). The ISC rate is proportional to the overlap between $|\chi_0\rangle$ (the ground vibrational state of $\Aone$) and $|\chi^\prime_{\nu_n}\rangle$ (an excited vibrational level of $\As$ that is separated from $\As$ by an energy spacing $\nu_n$).  We perform the sum over $n$ in order to define a vibrational overlap function
\begin{align}
F\hspace{-1.5 pt}\left(\Delta\right) &= \sum\limits_n |\langle \chi_0 | \chi^\prime_{\nu_n} \rangle|^2 \: \delta\hspace{-1 pt}\left(\nu_n - \Delta\right)
\nonumber \\
&= \overline{|\langle \chi_0 | \chi^\prime_\Delta \rangle|^2} \; \rho\hspace{-1.5 pt}\left(\Delta\right),
\end{align}
where $\rho\hspace{-1.5 pt}\left(\Delta\right)$ is the density of excited vibrational states that are resonant with $\Aone$ and the average is over all such states.

We substitute the vibrational overlap function, which encapsulates all relevant information about the quasi-continuum of $\As$ vibrational modes, into Eq. \ref{eq.A1 ISC 1} to find
\begin{equation}
\label{eq.A1 ISC}
\Gamma_{A_1} = 4\pi \hbar \, \lambda_\perp^2 \: F\hspace{-1.5 pt}\left(\Delta\right).
\end{equation}

\subsection{ISC Rate from $\Eonetwo$}
\label{sec.E12 ISC}

An analogous first-order ISC process is not responsible for the ISC transition from $\Eonetwo$; $\Eonetwo$ cannot decay to $\As$ because there is no SO coupling between these states, and we expect decay to $\Es$ to be negligible because the $\As-\Es$ energy spacing (1190 meV \cite{Acosta2010a}) is large compared to the extent of the phonon sideband ($\sim500$ meV \cite{Davies1974,Kehayias2013}).  Instead, ISC decay from $\Eonetwo$ is the result of a second-order process wherein phonons of $E$ symmetry couple $\Eonetwo$ to $\Aone$ and $\Aone$ is SO-coupled to $\As$. We calculate the ISC rate from $\Eonetwo$ using second-order Fermi's golden rule and find
\begin{align}
\label{eq.Initial GEonetwo}
\Gamma_{E_{1,2}} = 2 \pi \hbar^3 \sum\limits_{n,p,k} & \left|\frac{\sqrt{2} \, \lambda_\perp \lambda_{p,k}}{\omega_k}\right|^2 \,
|\langle \chi_0 | \chi^\prime_{\nu_n} \rangle|^2 \nonumber \\
& \, \times \big[ \left(n_{p,k}+1\right) \delta\hspace{-1 pt}\left(\nu_n + \omega_k - \Delta\right) \nonumber \\
& \;\;\;\;\, + n_{p,k} \: \delta\hspace{-1 pt}\left(\nu_n - \omega_k - \Delta\right) \big],
\end{align}
where $\omega_k$ is the energy of a phonon of wavevector $k$.  The two distinct terms within the sum correspond to phonon emission and absorption.

We introduce the $k$-independent phonon energy $\omega$, over which we integrate to pull out the polarization-specific phonon spectral density
\begin{equation}
\label{eq.Spectral Density Definition}
J_p\hspace{-1.5pt}\left(\omega\right)=\frac{\pi\hbar}{2}\sum\limits_k \lambda_{p,k}^2\,\delta\hspace{-1 pt}\left(\omega-\omega_k\right).
\end{equation}
In the linear dispersion regime, where the wavelength of acoustic phonons is much larger than the lattice spacing, the coupling strength for interactions mediated by deformations of the lattice is given by the standard deformation potential ($\lambda_{p,k}\propto\sqrt{\omega_k}$ \cite{Weiss2008}) and the phonon density of states is described by the Debye model ($\mathrm{DOS}\propto\omega^2$).  The total spectral density is therefore
\begin{equation}
\label{eq.Spectral Density}
J\hspace{-1.5pt}\left(\omega\right)=J_p\hspace{-1.5pt}\left(\omega\right)=\eta \; \omega^3,
\end{equation}
where $\eta$ parameterizes the coupling strength between the states of the $\Et$ manifold and $E$-symmetric acoustic phonons.

We insert this spectral density and the appropriate vibrational overlap functions to find
\begin{align}
\label{eq.E12 ISC with n}
\Gamma_{E_{1,2}} = 8 \: \hbar^2 \lambda_\perp^2 \eta \int\limits_0^\Omega \: \omega \:
\big\{ &\left[n\left(\omega\right)+1\right] F\hspace{-1.5 pt}\left(\Delta - \omega\right) \nonumber \\
& \hspace{-3 pt} + n\left(\omega\right) \: F\hspace{-1.5 pt}\left(\Delta + \omega\right) \big\} \: \mathrm{d}\omega,
\end{align}
where
\begin{equation}
n\left(\omega\right) = \frac{1}{ e^{\omega/k_BT} - 1}
\end{equation}
is the thermal occupation of a phonon mode of energy $\omega$.  We assume a cutoff energy $\Omega$ for acoustic phonons.

\begin{figure}[h]
\begin{center}
\includegraphics[width=\columnwidth]{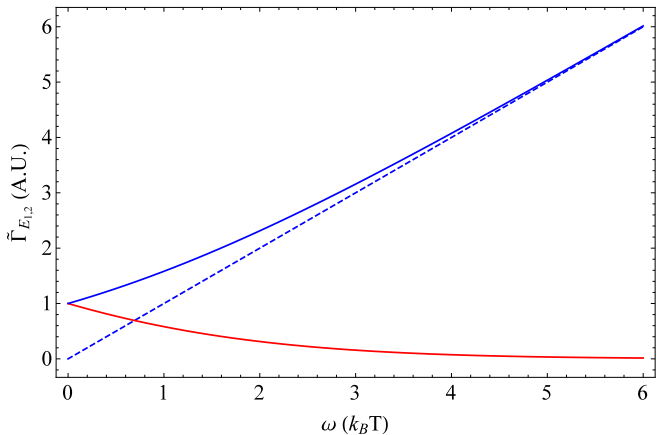}
\caption{ISC rate $\tilde{\Gamma}_{E_{1,2}}$ from $\Eonetwo$ as a function of the energy of the mediating $E$-symmetric phonon.  Contributions due to phonon absorption (red), stimulated and spontaneous emission (blue), and spontaneous emission only (blue dashed) are shown.  At $T=5$ K, $k_BT=2\pi\hbar\times104\,\mathrm{GHz}=0.43$ meV.  We assume that the vibrational overlap function $F\hspace{-1.5 pt}\left(\Delta\pm\omega\right)$ is flat range for the range of $\omega$ (2.6 meV) shown.}
\label{fig.GE12 vs omega}
\end{center}
\end{figure}

To illustrate the range of phonon energies that contribute to $\GEonetwo$, we define a rate $\tilde{\Gamma}_{E_{1,2}}\hspace{-2 pt}\left(\omega\right)$ that is mediated only by $E$-symmetric phonons of energy $\omega$, such that $\Gamma_{E_{1,2}}=\int_0^\Omega \tilde{\Gamma}_{E_{1,2}}\hspace{-2 pt}\left(\omega\right) \mathrm{d}\omega$.  The contributions to $\tilde{\Gamma}_{E_{1,2}}\hspace{-2 pt}\left(\omega\right)$ due to phonon emission and absorption are shown in Fig. \ref{fig.GE12 vs omega}.  The dominant contribution to $\GEonetwo$ comes from high-energy phonon modes, whose thermal occupations are negligible at $T<26$ K.  This conclusion is valid for all values of $\Delta$ because the vibrational overlap function $F\hspace{-1.5 pt}\left(\Delta\pm\omega\right)$ changes slowly on the scale of $k_BT\lesssim5$ meV.  We may therefore neglect thermal occupation of the mediating phonon modes and take the low-temperature limit of $\GEonetwo$ when working at cryogenic temperatures, assertions that we will justify in Sec. \ref{sec.Quant ISC from E12}.

In the low-temperature limit, the ISC rate from $\Eonetwo$ is given by
\begin{equation}
\label{eq.E12 ISC}
\Gamma_{E_{1,2}} = \frac{2}{\pi} \, \hbar \, \eta \: \Gamma_{A_1} \hspace{-13 pt} \int\limits_0^{\min\left(\Delta,\Omega\right)} \hspace{-11 pt} \omega \;\,
\frac{F\hspace{-1.5 pt}\left(\Delta-\omega\right)}{F\hspace{-1.5 pt}\left(\Delta\right)}
\: \mathrm{d}\omega.
\end{equation}
The range of phonon modes that contribute to $\GEonetwo$ is bounded either by the cutoff energy $\Omega$, or by the fact that $F\hspace{-1.5 pt}\left(\Delta-\omega\right)=0$ for $\omega\geq\Delta$ at low temperature [see Sec. \ref{sec.Thermal Modification of Phononic Band}].

If second-order ISC processes that use $\Es$ as intermediate states are taken into account, as described in the appendix, then $\GEonetwo$ is modified to
\begin{align}
\label{eq.E12 ISC with 1E12}
\Gamma_{E_{1,2}} = \frac{2}{\pi} \, \hbar \, \eta \: \Gamma_{A_1} \hspace{-13 pt} \int\limits_0^{\min\left(\Delta,\Omega\right)} \hspace{-10 pt} & \omega^3 \: \left(\frac{1}{\omega}-\frac{2}{\Delta+\Delta^\prime}\right)^2 \nonumber \\
& \times \frac{F\hspace{-1.5 pt}\left(\Delta-\omega\right)}{F\hspace{-1.5 pt}\left(\Delta\right)} \: \mathrm{d}\omega.
\end{align}
We note that this secondary ISC process only contributes significantly to the total ISC rate from $\Eonetwo$ because it interferes coherently with the primary ISC process, which uses $\Aone$ as the intermediate state.

\subsection{Phonon-Induced Mixing Rate}
\label{sec.Explicit Calculation of GMix}

We now explicitly calculate the phonon-induced mixing rates between the states in the $\Et$ manifold.  Doing so will enable us to extract the value of $\eta$, which parameterizes the NV-phonon coupling strength, from a temperature-dependent measurement of the $\Ex-\Ey$ mixing rate \cite{Goldman2014}.

We use a technique similar to that used to calculate the ISC rate from $\Eonetwo$ in the previous section.  We begin with the mixing rate between $\Ex$ and $\Ey$.  The dominant contribution to the mixing rate in the low-strain regime ($\Delta_{xy}=\mathcal{E}_{E_x}-\mathcal{E}_{E_y}=2\pi\hbar\times3.9$ GHz for the experiment in question \cite{Goldman2014}) is from a two-phonon Raman processes wherein one phonon is emitted and another is absorbed \cite{Fu2009}.  Following Ref. \onlinecite{Fu2009}, we use second-order Fermi's golden rule to find that this mixing rate is
\begin{align}
\label{eq.GMix explicit}
\Gamma_\mathrm{Mix} = \frac{32\hbar}{\pi} \hspace{-2pt} \int\limits_0^\infty & n(\omega)\left[n(\omega+\Delta_{xy})+1\right] \nonumber \\
&\times \bigg[\frac{J_{1}\hspace{-2pt}\left(\omega+\Delta_{xy}\right)J_{2}\hspace{-2pt}\left(\omega\right)}{(\omega+\Delta_{xy})^2} \nonumber \\
&\;\;\;\;\;+ \frac{J_{1}\hspace{-2pt}\left(\omega\right)J_{2}\hspace{-2pt}\left(\omega+\Delta_{xy}\right)}{\omega^2}\bigg] \mathrm{d}\omega.
\end{align}

\begin{figure}[h]
\begin{center}
\includegraphics[width=\columnwidth]{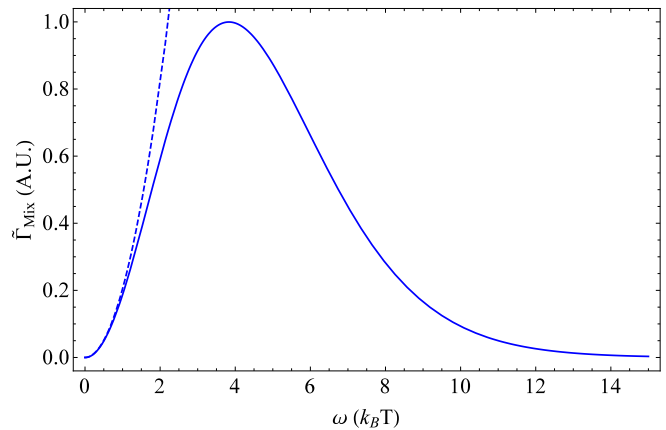}
\caption{Mixing rate $\tilde{\Gamma}_\mathrm{Mix}$ as a function of the energy of the lower-energy mediating $E$-symmetric phonons.  At $T=5$ K, $k_BT=2\pi\hbar\times104\,\mathrm{GHz}=0.43$ meV.  The dashed line shows $\omega^2$, properly scaled, for comparison.}
\label{fig.GMix vs omega}
\end{center}
\end{figure}

We substitute the expression for the spectral density $J_p\hspace{-1.5pt}\left(\omega\right)$ in the acoustic limit (Eq. \ref{eq.Spectral Density}).  We keep only the highest-order term in $\omega/\Delta_{xy}$ because the integrand in Eq. \ref{eq.GMix explicit} is only appreciable for $\omega \sim k_BT \gg \Delta_{xy}$ and, as in the previous section, we define an mixing rate
\begin{equation}
\tilde{\Gamma}_\mathrm{Mix}\hspace{-2 pt}\left(\omega\right) = \frac{64}{\pi} \, \hbar \: \eta^2 \, \omega^4 \, n(\omega)\left[\,n(\omega+\Delta_{xy})+1\,\right]
\end{equation}
that is mediated only by phonons of energies $\omega$ and $\omega+\Delta_{xy}$.  As is shown in Fig. \ref{fig.GMix vs omega}, contributions to the total mixing rate $\Gamma_\mathrm{Mix}=\int_0^\infty \tilde{\Gamma}_\mathrm{Mix}\hspace{-2 pt}\left(\omega\right) \mathrm{d}\omega$ are dominated by phonon modes with energy of order $k_BT$.  Higher-energy phonons have significantly larger spectral densities, but the emission-absorption Raman process requires that the phonon mode of energy $\omega$ have non-negligible thermal occupation.  Unlike with the ISC rate from $\Eonetwo$, which is mediated mainly by high-energy phonons, the contributions to the mixing rate from phonon modes with energies larger than approximately 20 meV are exponentially suppressed at cryogenic temperatures, so we need not impose a cutoff energy for the available phonon modes.

We therefore find the total mixing rate
\begin{equation}
\label{eq.Ex-Ey Mixing Rate}
\Gamma_\mathrm{Mix} = \frac{64}{\pi} \, \hbar \: \alpha \, \eta^2 k_B^5 \, T^5.
\end{equation}
The numeric constant $\alpha$ is given by the integral
\begin{equation}
\label{eq.alpha}
\alpha = \int_0^{\infty} \frac{1}{e^x-1} \left(\frac{1}{e^{x+x_\Delta}-1}+1\right) x^4 \, \mathrm{d}x,
\end{equation}
where $x=\omega/k_BT$ and $x_\Delta=\Delta_{xy}/k_BT$.

The mixing rates due to other processes are negligible.  There are two alternate two-phonon processes, wherein both phonons are either absorbed or emitted, but they do not conserve energy and contribute negligibly, respectively.  There is a direct one-phonon emission process, for which we calculate a transition rate
\begin{equation}
\label{eq.One-phonon mixing rate}
\Gamma_\mathrm{Mix}^{(1-\mathrm{ph})} = 4 \, \eta\left[\,n(\Delta_{xy})+1\,\right]\Delta_{xy}^3\approx 4 \, \eta \, k_B \, \Delta_{xy}^2 \, T,
\end{equation}
where the approximation is true for $\Delta_{xy}\ll k_BT$.  This one-phonon rate is negligible when the strain splitting $\Delta_{xy}$ is small.  For example, we calculate that a strain splitting of at least $\Delta_{xy}/2\pi\hbar = 18$ GHz is required to produce a measurable ($\geq0.5$ MHz) one-phonon mixing rate at 5 K, whereas this analysis assumes $\Delta_{xy}/2\pi\hbar = 3.9$ GHz to match Ref. \onlinecite{Goldman2014}.  We note that a mixing rate that scales as $T$ rather than $T^5$ would be more difficult to suppress using standard liquid helium cryogenic techniques with $T\sim5$ K.  This consideration may explain why a previous measurement of phonon-induced mixing \cite{Fu2009} found that the $\Ex-\Ey$ mixing rate was increased for NV centers with higher strain splittings (44 to 81 GHz, compared with 8 and 9 GHz) and did not appear to asymptote to a constant value at low temperatures to the same degree.  The effect of the one-phonon mixing process may be even more substantial for NV centers, such as those formed by nitrogen implantation or placed inside nanofabricated structures, where damage to the local crystalline structure may induce a large strain splitting.

We now calculate the phonon-induced mixing rate between $\Aone$ and $\Atwo$.  This calculation is similar to that of the $\Ex-\Ey$ mixing rate.  Instead of coupling and splitting the energies of $\Ex$ and $\Ey$, phonons of $E$ symmetry couple $\Aone$ and $\Atwo$ with $\Eone$ and $\Etwo$.  The fact that the intermediate electronic state ($\Eone$ or $\Etwo$) is not degenerate with the initial or final state ($\Aone$ or $\Atwo$), gives rise to a resonance condition when the energy of the emitted phonon is equal to the splitting between the initial and intermediate electronic states.  This resonance condition, however, is irrelevant because the mixing process is dominated by phonons with energy $\omega\approx 4 k_B T \approx 2\pi\hbar\times400$ GHz, as shown in Fig. \ref{fig.GMix vs omega}.  Because the typical phonon energy is large compared to the 10 GHz splitting between $\Aone$ or $\Atwo$ and $\Eonetwo$, the mixing rate between $\Aone$ and $\Atwo$ is unaffected by the detuning of the intermediate state and is also given by Eq. \ref{eq.Ex-Ey Mixing Rate}.  The mixing rate between $\Eone$ and $\Etwo$ is also given by Eq. \ref{eq.Ex-Ey Mixing Rate} for the same reasons, although mixing between these two states cannot be observed directly using the techniques employed in Ref. \onlinecite{Goldman2014} because the ISC rates from both states are the same.

Further, both $\Aone$ and $\Atwo$ are coupled to both $\Eone$ and $\Etwo$ by a one-phonon process, whose rate in all four cases is given by Eq. \ref{eq.One-phonon mixing rate} with $\Delta_{xy}$ replaced by the appropriate energy splitting.  For the reverse process, from $\Eone$ and $\Etwo$ to $\Aone$ and $\Atwo$, we replace $\left[\,n(\Delta_{xy})+1\,\right]$ with $n(\Delta_{xy})$, but the result is the same in the limit $\Delta_{xy}\ll k_BT$.  Thus, the cyclicity of the nominally closed $|\pm1\rangle-\Atwo$ $\Lambda$ system, like that of the $|0\rangle-\Ex$ cycling transition, may be degraded in NV centers with high strain splittings.

\section{Comparison to Measurements}
\label{sec.Comparison to Measurements}

\subsection{ISC Rate from $\Aone$}
\label{sec.Quant ISC from A1}

\begin{figure}[h]
\begin{center}
\includegraphics[width=\columnwidth]{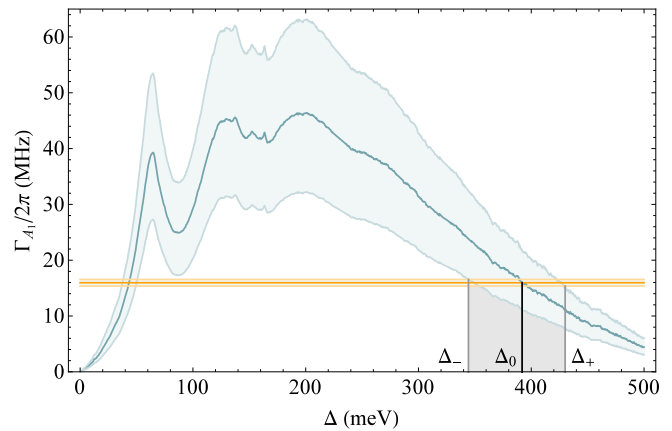}
\caption{ISC rate from $\Aone$.  The values of $\GAone$ calculated using Eq. \ref{eq.A1 ISC} (blue) and measured in Ref. \onlinecite{Goldman2014} (orange) are shown, with confidence intervals given by the uncertainty bounds on $\lambda_\perp$ described in the text and the $95\%$ confidence interval given for the measurement.  The predicted range of $\Delta$, which is defined by the intersection of the two curves, is shown in black.  The points of intersection below 148 meV are excluded by the measured $\GEonetwo/\GAone$ ratio, as explained in Sec. \ref{sec.Quant ISC from E12}.}
\label{fig.A1 ISC vs data}
\end{center}
\end{figure}

We now compare the results of the preceding calculations to the measured state-selective ISC rates \cite{Goldman2014}, with the goal of extracting the previously unknown energy spacing between the spin-singlet and spin-triplet levels.  We first compare the calculated ISC rate from $\Aone$, which depends on the transverse SO coupling rate $\lambda_\perp$ and the vibrational overlap function $F\hspace{-1.5 pt}\left(\omega\right)$, with the measured $\GAone/2\pi=16.0\pm0.6$ MHz.

Most of the uncertainty in our theoretical prediction of $\GAone$ is the result of the lack of precision with which $\lambda_\perp$ is known. The axial SO coupling rate $\lambda_{||}=5.33\pm0.03$ GHz has been measured precisely through spectroscopy of the $\Et$ manifold \cite{Bassett2014}, but $\lambda_\perp$ cannot be determined through similar methods. An approximate theoretical argument implies that $\lambda_\perp\sim\lambda_{||}$. Due to the similar mass and charge of nitrogen and carbon atoms, the $C_{3v}$ symmetric structure of the NV center is only a small departure from the $T_d$ symmetric structure of a vacancy in the diamond lattice. Given that $\lambda_\perp=\lambda_{||}$ precisely in $T_d$ symmetry, it is expected that $\lambda_\perp\sim\lambda_\parallel$ in the slightly perturbed symmetry of the NV center \cite{Manson2010}. This argument is supported by preliminary \textit{ab initio} calculations that estimate the ratio $\lambda_\perp/\lambda_{||}\sim1.15-1.33$ \cite{Doherty2012,Maze2011}. To reflect the uncertainty in the value of $\lambda_\perp$, we select a confidence band of $\lambda_\perp/\lambda_{||}=1.2\pm0.2$.

We extract the vibrational overlap function $F\hspace{-1.5 pt}\left(\omega\right)$ from spectroscopy of the $\Et\rightarrow\At$ emission PSB conducted at 4 K \cite{Kehayias2013}.  As we describe in Sec. \ref{sec.Thermal Modification of Phononic Band}, $F\hspace{-1.5 pt}\left(\omega\right)$ is, in principle, temperature-dependent.  However, the dominant and lowest-energy feature in the one-phonon spectrum extracted from the $\Et\rightarrow\At$ PSB occurs at $64\,\mathrm{meV}=k_B\times 118$ K \cite{Kehayias2013}.  Because phonons with such large energies are negligibly occupied at temperatures below 26 K, we assume that $F\hspace{-1.5 pt}\left(\omega\right)$ is effectively temperature-independent for our analysis of the low-temperature ISC rate.

In Fig. \ref{fig.A1 ISC vs data}, we compare the measured and predicted values of $\GAone$ in order to extract $\Delta$, the previously unknown $\Aone-\As$ energy splitting.  This analysis confines $\Delta$ to two regions: around 43 meV, and from $\Delta_-=344$ meV to $\Delta_+=430$ meV with a central value of $\Delta_0=392$ meV.  We exclude values of $\Delta$ below 148 meV because the predicted ISC rate from $\Eonetwo$ would be significantly lower than the observed rate, as explained in Sec. \ref{sec.Quant ISC from E12}.  The uncertainty in $\Delta$ is dominated by the uncertainty in $\lambda_\perp$, so a precise calculation of the $\lambda_\perp/\lambda_{||}$ ratio could narrow the bounds on $\Delta$ considerably.


\subsection{ISC Rate from $\Eonetwo$}
\label{sec.Quant ISC from E12}

\begin{figure}[h]
\begin{center}
\includegraphics[width=\columnwidth]{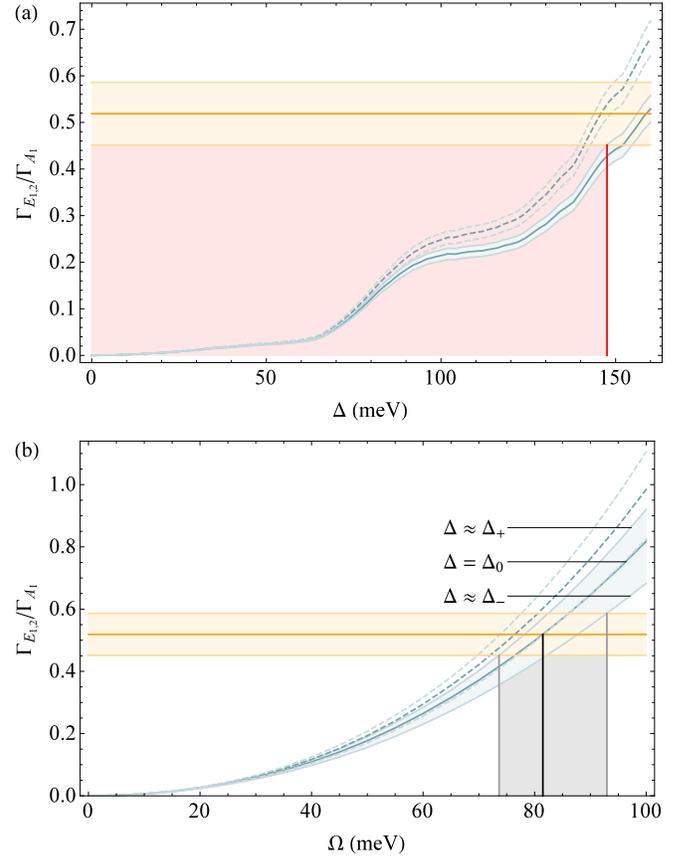}
\caption{Ratio of the ISC rates from $\Eonetwo$ and $\Aone$.  The values of $\GEonetwo/\GAone$ calculated using Eq. \ref{eq.E12A1 Ratio} (blue) and measured in Ref. \onlinecite{Goldman2014} (orange, with $95\%$ confidence interval) are shown.  The dashed plots represent the results obtained when the ISC process that uses $\Es$ as intermediate states is neglected (see appendix).  In (a), we assume no acoustic phonon cutoff ($\Omega\rightarrow\infty$), so the blue plot represents an upper bound on $\GEonetwo/\GAone$.  The calculated value's uncertainty is due to the uncertainty in $\eta$.  This comparison excludes $\Delta<148$ meV, as indicated by the red region.  In (b), we use a range of $\Delta$ that corresponds approximately \cite{DeltaRangeFootnote} to the range shown in Fig. \ref{fig.A1 ISC vs data}, and we vary the acoustic cutoff energy $\Omega$.  The grey region represents the predicted range of $\Omega$ \cite{OmegaRangeFootnote}.}
\label{fig.ISC ratio vs data}
\end{center}
\end{figure}

We also compare the measured and predicted ratios of the ISC rates from $\Eonetwo$ and $\Aone$.  This ratio, which is given by
\begin{align}
\label{eq.E12A1 Ratio}
\Gamma_{E_{1,2}}/\Gamma_{A_1} = \frac{2}{\pi} \, \hbar \, \eta \hspace{-13 pt} \int\limits_0^{\min\left(\Delta,\Omega\right)} \hspace{-10 pt} & \omega^3 \: \left(\frac{1}{\omega}-\frac{2}{\Delta+\Delta^\prime}\right)^2 \nonumber \\
& \times \frac{F\hspace{-1.5 pt}\left(\Delta-\omega\right)}{F\hspace{-1.5 pt}\left(\Delta\right)} \: \mathrm{d}\omega.
\end{align}
has the advantage of being insensitive to the uncertainty in $\lambda_\perp$, which limits the precision of our determination of $\Delta$.  It is, however, sensitive to uncertainty concerning the range of acoustic $E$-symmetric phonon modes that contribute to $\GEonetwo$.  We use the expression for $\GEonetwo$ derived in the low-temperature limit (Eq. \ref{eq.E12 ISC}), which we will justify at the end of this section.  We use the value of $\eta=2\pi\times\left(44.0\pm2.4\right)\,\mathrm{MHz}\:\mathrm{meV}^{-3}$ extracted from a measurement of $\GMix$ as a function of temperature \cite{Goldman2014}.

We analyze the $\GEonetwo/\GAone$ ratio in both of the regions identified by our analysis of $\GAone$: $\Delta\approx43\,\mathrm{meV}$ and $344\,\mathrm{meV}\leq\Delta\leq430\,\mathrm{meV}$.  In the first region, shown in \crefformat{figure}{Fig.~#2#1{(a)}#3}\cref{fig.ISC ratio vs data}, the predicted ratio is significantly lower than the measured ratio, even when we assume no acoustic cutoff energy ($\Omega\rightarrow\infty$).  This inconsistency excludes values of $\Delta$ up to 148 meV, which eliminates the $\Delta\approx43\,\mathrm{meV}$ region entirely.  In the second region, we do not assume an infinite cutoff energy, but instead find the cutoff energy that would make the predicted ratio consistent with measurement, as shown in \crefformat{figure}{Fig.~#2#1{(b)}#3}\cref{fig.ISC ratio vs data}. Our analysis predicts an acoustic phonon cutoff energy $\Omega$ between 74 and 93 meV.  This range coincides with a sharp and significant decline in the phonon spectral density extracted from the absorption PSB of the $\At\rightarrow\Et$ optical transition \cite{TripletEPSBFootnote,Davies1974}.  This agreement implies that the range of the predicted phonon cutoff energy is physically sensible, despite being near the upper limit of the acoustic phonon regime \cite{Pavone1993,Warren1967}.  Our approximate expression for the phonon spectral density (Eq. \ref{eq.Spectral Density}) is derived in the acoustic limit, but any correction that should be made to this approximation is swept into the phenomenological cutoff $\Omega$.  Hence, we conclude that the second region $344\,\mathrm{meV}\leq\Delta\leq430\,\mathrm{meV}$ is fully consistent with the observed ISC rates.

\begin{figure}[h]
\begin{center}
\includegraphics[width=\columnwidth]{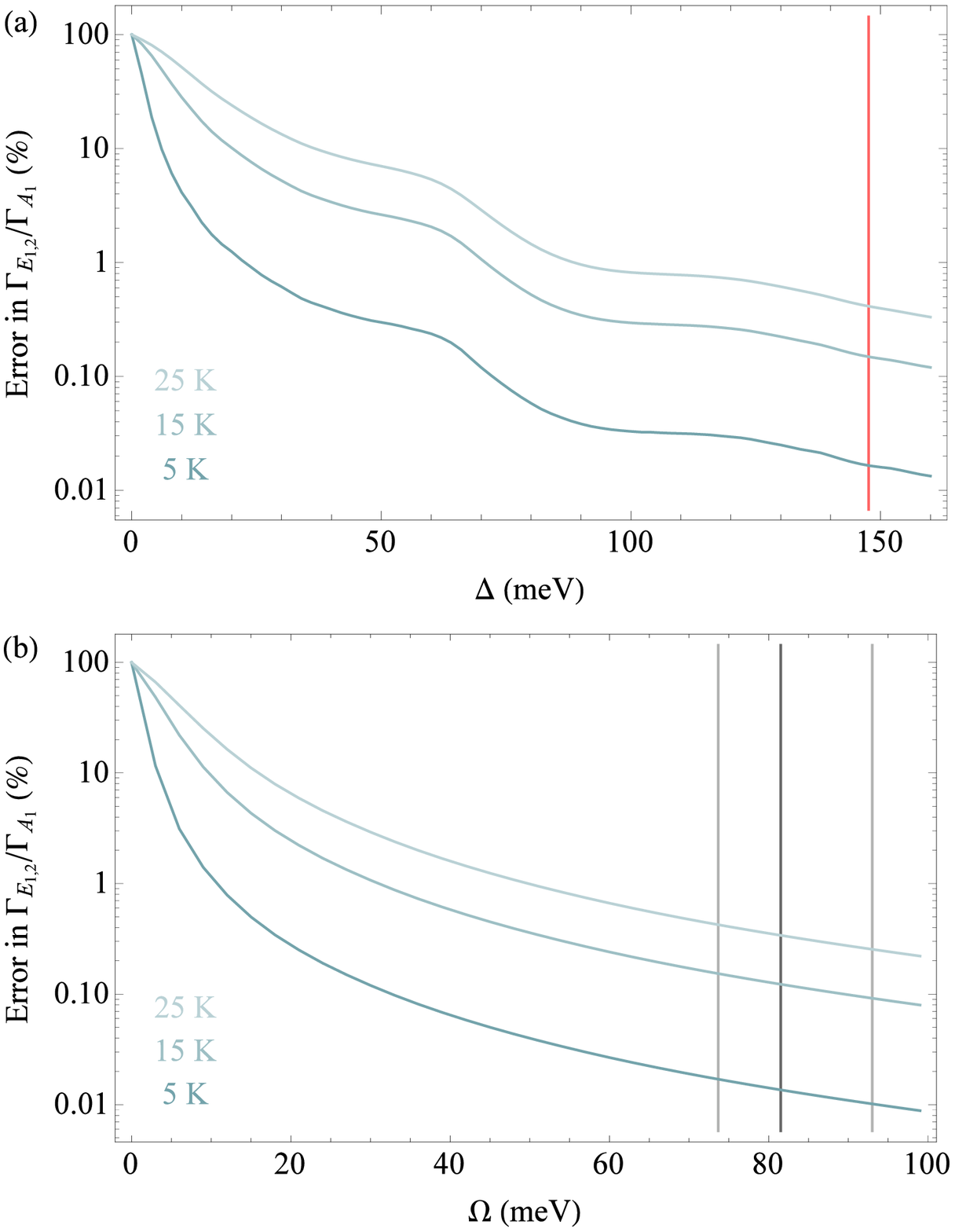}
\caption{Error in $\GEonetwo/\GAone$ ratio due to assumption of the low-temperature limit.  The theoretical values of $\GEonetwo/\GAone$ shown in Fig. \ref{fig.ISC ratio vs data} were calculated using Eq. \ref{eq.E12A1 Ratio}, which neglects thermal occupation of the mediating phonon modes.  We calculate the error due to working in the low-temperature limit by using the temperature-dependent expression for $\GEonetwo$ (Eq. \ref{eq.E12 ISC with n}) rather than the temperature-independent expression (Eq. \ref{eq.E12 ISC}).  We perform this calculation for the cases where (a) $\Omega$ is infinite and $\Delta$ is varied, and where (b) $\Delta$ is given by the values extracted from our analysis of $\GAone$ and $\Omega$ is varied.  These are the same cases shown in the corresponding subfigures of Fig. \ref{fig.ISC ratio vs data}.  The results extracted from Fig. \ref{fig.ISC ratio vs data}, the minimum allowed value of $\Delta$ in (a) and the expected values of $\Omega$ in (b), are shown to emphasize that the conclusions we draw using the low-temperature limit are valid.}
\label{fig.Low-T error}
\end{center}
\end{figure}

We now support the assumption made in Sec. \ref{sec.E12 ISC} that we may neglect thermal occupation of the $E$-symmetric phonon modes that mediate the ISC from $\Eonetwo$.  In Fig. \ref{fig.Low-T error}, we calculate the error due to neglecting the contribution to $\Eonetwo$ due to stimulated emission and absorption of phonons.  While this assumption would not be valid if either $\Delta$ or $\Omega$ were small, we see that the resulting error is less than $1\%$ for the values of $\Delta$ and $\Omega$ extracted from our analyses of $\GAone$ and $\GEonetwo/\GAone$.

\section{Extension to High Temperatures}
\label{sec.Extension of Model to High Temperatures}

\subsection{Method}
\label{sec.Thermal Modification of Phononic Band}

We scale our model up to higher temperatures to compare the predicted lifetimes of the $\Et$ states with $\mso$ to measurements conducted at temperatures between 295 K and 700 K \cite{Toyli2012,Batalov2008,Robledo2011}.  In Secs. \ref{sec.A1 ISC} and \ref{sec.E12 ISC}, we calculated the ISC rates from $\Aone$ and $\Eonetwo$ in the low-temperature limit.

Extension of the ISC model to higher temperatures requires three modifications to the ISC calculation: $\left(1\right)$ Because of phonon-induced orbital averaging \cite{Rogers2009}, which is significant even at $T\sim20$ K \cite{Goldman2014}, the observed ISC rate will be the average ISC rate from all $\Et$ states with $\mso$. $\left(2\right)$ The $E$-symmetric phonon modes that mediate the ISC transition from $\Eonetwo$ have non-negligible thermal occupation.  We must therefore consider ISC contributions due to stimulated and spontaneous emission into these modes, as well as absorption from these modes. $\left(3\right)$ The $A_1$-symmetric phonon modes that are primarily responsible for shifting the lattice from its $\Et$ equilibrium configuration to its $\As$ equilibrium configuration have non-negligible thermal occupation, so the vibrational overlap function $F\hspace{-1.5 pt}\left(\omega\right)$ becomes broader and flatter at higher temperatures.

To address modification $\left(1\right)$, we calculate the orbitally averaged ISC rate
\begin{equation}
\label{eq.GISC Total}
\Gamma_\mathrm{ISC} = \frac{1}{4}\left(\GAone+2\:\GEonetwo\right)
\end{equation}
where $\GAone$ is given by Eq. \ref{eq.A1 ISC}.

To address modification $\left(2\right)$, we use $\GEonetwo$ as given by Eq. \ref{eq.E12 ISC with n}, which includes contributions due to both phonon absorption and emission, instead of Eq. \ref{eq.E12 ISC}, which is calculated in the low-temperature limit.

\begin{figure}[h]
\begin{center}
\includegraphics[width=\columnwidth]{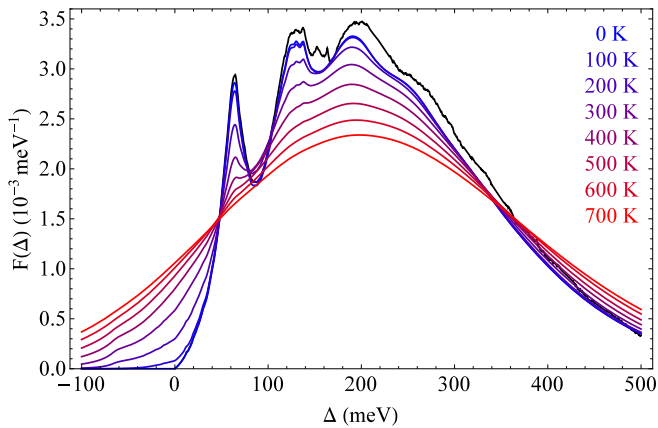}
\caption{Calculated temperature-dependent vibrational overlap functions $F\hspace{-2 pt}\left(\omega,T\right)$.  The measured low-temperature vibrational overlap function $F\hspace{-2 pt}\left(\omega\right)$ is shown in black \cite{Kehayias2013}.}
\label{fig.High-T Phononic Bands}
\end{center}
\end{figure}

To address modification $\left(3\right)$, we calculate the temperature-dependent vibrational overlap function $F\hspace{-2 pt}\left(\omega,T\right)$ using the procedure given in Ref. \onlinecite{Davies1974}.  We first extract the low-temperature one-phonon spectral density $f\hspace{-2 pt}\left(\omega\right)$ by numerically deconvolving the low-temperature phonon sideband $F\hspace{-2 pt}\left(\omega\right)$, which was used in Secs. \ref{sec.A1 ISC} and \ref{sec.E12 ISC} to calculate the low-temperature ISC rates from $\Aone$ and $\Eonetwo$.  The resulting $f\hspace{-2 pt}\left(\omega\right)$ is shown in red in Fig. 4 of Ref. \onlinecite{Kehayias2013}, whose measurement of $F\hspace{-2 pt}\left(\omega\right)$ we use throughout our analysis.

The temperature-dependent one-phonon vibrational overlap function is given by
\begin{equation}
F_1\hspace{-2 pt}\left(\omega,T\right) = \begin{cases}
\left[n\hspace{-1.5 pt}\left(\omega,T\right)+1\right] \: f\hspace{-2 pt}\left(\omega\right) & \text{if} \: \omega \geq 0 \\
n\hspace{-1.5 pt}\left(\omega,T\right) \: f\hspace{-2 pt}\left(-\omega\right) & \text{if} \: \omega < 0
\end{cases}.
\end{equation}
Temperature-dependent multi-phonon vibrational overlap functions can be calculated recursively using
\begin{align}
F_i\hspace{-2 pt}\left(\omega,T\right) &= F_{i-1}\hspace{-2 pt}\left(\omega,T\right) \otimes F_1\hspace{-2 pt}\left(\omega,T\right) \nonumber \\
&= \int\limits_{-\infty}^\infty F_{i-1}\hspace{-2 pt}\left(\omega-\omega^\prime,T\right) \: F_1\hspace{-2 pt}\left(\omega^\prime,T\right) \:
\mathrm{d}\omega^\prime
\end{align}
and then summed to find the total temperature-dependent vibrational overlap function
\begin{equation}
\label{eq.T-dependent F}
F\hspace{-2 pt}\left(\omega,T\right) = e^{-S} \sum\limits_{i=1}^\infty \frac{S^i}{i!} \, F_i\hspace{-2 pt}\left(\omega,T\right).
\end{equation}

Here,
\begin{equation}
S = S_0 \int\limits_0^\Omega \left[ 2 \: n\hspace{-1.5 pt}\left(\omega^\prime,T\right) +1 \right] \: f\hspace{-2 pt}\left(\omega^\prime\right) \:
\mathrm{d}\omega^\prime
\end{equation}
is the temperature-dependent Huang-Rhys factor, where $S_0=3.49$ is the low-temperature Huang-Rhys factor for the $\Et\rightarrow\At$ transition \cite{Kehayias2013}.

In Fig. \ref{fig.High-T Phononic Bands}, we plot the calculated $F\hspace{-2 pt}\left(\Delta,T\right)$ for temperatures between 0 and 700 K.  We note a slight discrepancy between the measured low-temperature $F\hspace{-2 pt}\left(\omega\right)$ and the calculated $F\hspace{-2 pt}\left(\omega,T\right)$ at 0 K.  There is a degree of imprecision inherent in this method of generating $F\hspace{-2 pt}\left(\omega,T\right)$, but it should be sufficient to calculate the temperature trend of the theoretical ISC rates.

\subsection{Results}

\begin{figure}[h]
\begin{center}
\includegraphics[width=\columnwidth]{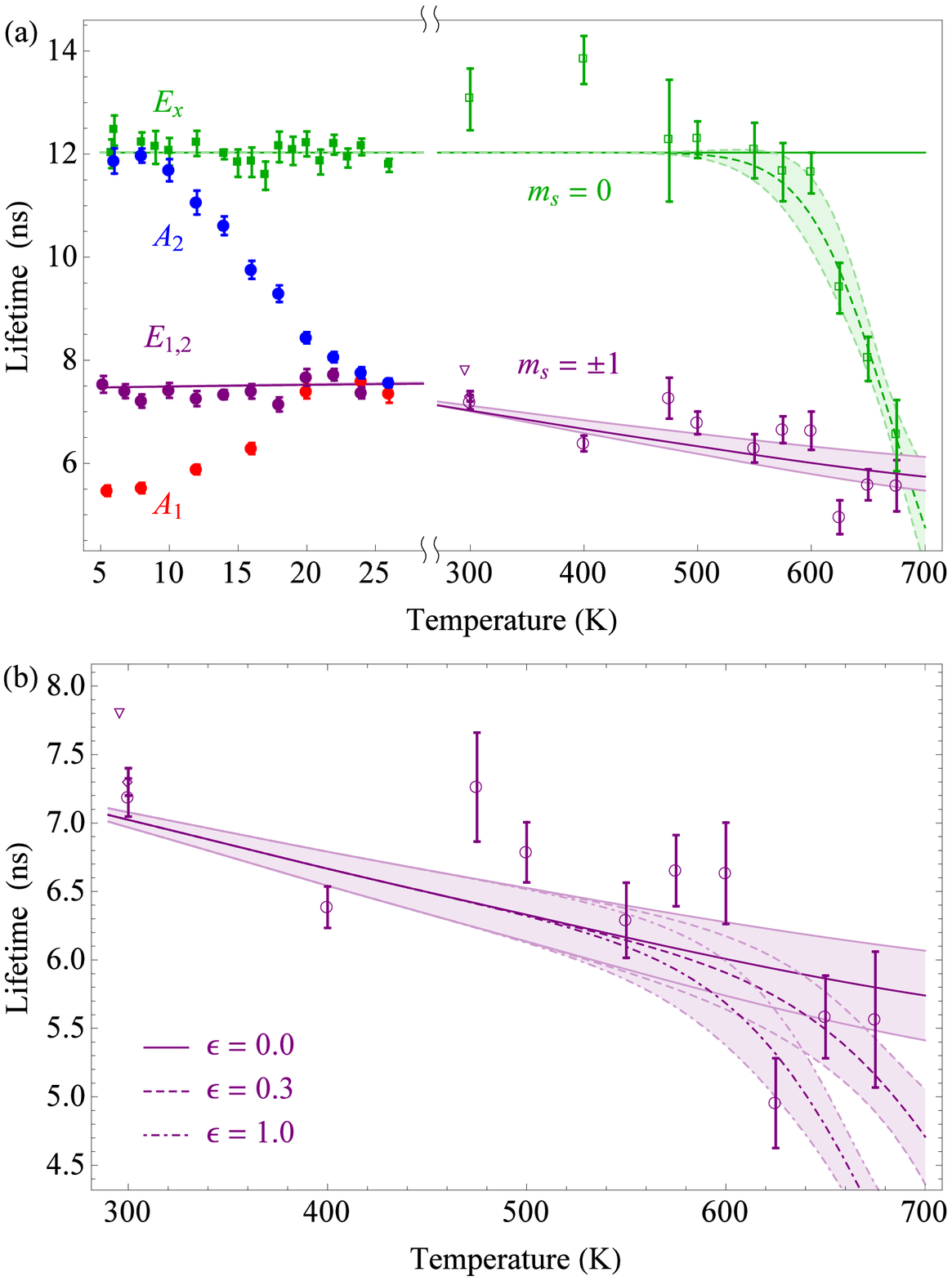}
\caption{Comparison of the predicted high-temperature fluorescence lifetimes with measured lifetimes.  The data from 5 K to 26 K are taken from Ref. \onlinecite{Goldman2014} and the data from 295 K to 700 K are taken from Refs. \onlinecite{Toyli2012} ($\Box$ and $\ocircle$), \onlinecite{Robledo2011} ($\diamond$), and \onlinecite{Batalov2008} ($\triangledown$).  In (a), the solid green line is the lifetime of the $\msz$ states predicted by our model, and the dashed green line includes a fit to the Mott-Seitz model with $95\%$ confidence interval, as described in the text.  This decay mechanism is not included in the predicted lifetime of the $\mso$ states.  In (b), we plot the predicted lifetime of the $\mso$ states with varying degrees of coupling to the high-temperature decay mechanism.}
\label{fig.High-T Results}
\end{center}
\end{figure}

We now numerically calculate the orbitally averaged ISC rate for the $\mso$ states (Eq. \ref{eq.GISC Total}) using the temperature-dependent expressions for $\GEonetwo$ (Eq. \ref{eq.E12 ISC with n}) and $F\hspace{-2 pt}\left(\omega,T\right)$ (Eq. \ref{eq.T-dependent F}, plotted in Fig. \ref{fig.High-T Phononic Bands}).  We convert this ISC rate to the observed fluorescence lifetime using
\begin{equation}
\label{eq.Lifetime from ISC}
\tau=\frac{1}{\GRad+\GISC},
\end{equation}
where $\GRad = 2\pi\times\left(13.2\pm0.5\right)$ MHz is the radiative decay rate of states in the $\Et$ manifold \cite{Goldman2014}.  We use as numeric inputs the range of $\Delta$ and the associated values of $\Omega$, extracted from the analyses shown in Figs. \ref{fig.A1 ISC vs data} and \crefformat{figure}{#2#1{(b)}#3}\cref{fig.ISC ratio vs data}, that produce the measured $\GEonetwo/\GAone$ ratio at low temperature.

We plot the results of this calculation (purple), as well as the predicted temperature-independent lifetime of the $\msz$ states (solid green), in \crefformat{figure}{Fig.~#2#1{(a)}#3}\cref{fig.High-T Results}.  We observe that the predicted lifetimes are consistent with experimental observations \cite{Toyli2012,Batalov2008,Robledo2011} for temperatures below 600 K.  Above 600 K, it is clear that a temperature-dependent decay process for the $\msz$ states switches on. This process may also shorten the average lifetime of the $\mso$ states above 600 K, but the evidence for this assertion is not conclusive. This process has been previously attributed to a multi-phonon nonradiative relaxation from $\Et$ to $\As$ and described by a Mott-Seitz model \cite{Toyli2012}. We fit the high-temperature lifetime of the $\msz$ states (dashed green) using the Mott-Seitz model and the $\Ex$ lifetime of 12.0 ns that has been measured at low temperature \cite{Goldman2014} to obtain the temperature-dependent decay rate $\Gamma_\mathrm{HT}$ from the $m_s=0$ states due to this high-temperature mechanism \footnote{In terms of the Mott-Seitz formula given in Ref. \onlinecite{Toyli2012}, the high-temperature decay rate is given by $\Gamma_\mathrm{HT}\left(T\right) = s\,\GRad \,e^{-\Delta E/k_BT}$, where $s$ is the frequency factor and $\Delta E$ is the energy barrier for the nonradiative process.  We find $\Delta E = 0.94\pm0.32$ eV and $s<5.8\times10^7$, which differ significantly from the values given in Ref. \onlinecite{Toyli2012}.  This disagreement is not unexpected, however, because we assume a low-temperature lifetime of 12.0 ns whereas Toyli et al. fit a low-temperature lifetime of $13.4\pm0.6$ ns.  Our fit therefore exhibits a sharper turn-on, resulting in a higher value of $\Delta E$; the frequency factor $s$, which sets the vertical scaling and is extremely sensitive to $\Delta E$, is correspondingly much larger.}.  To incorporate $\Gamma_\mathrm{HT}$ into the predicted lifetime of the $\mso$ states, we make the replacement $\GISC\rightarrow\GISC+\epsilon\,\Gamma_\mathrm{HT}$ in Eq. \ref{eq.Lifetime from ISC}, where $\epsilon$ parameterizes how strongly the high-temperature decay mechanism couples to the $\mso$ states relative to the $\msz$ states.  The predicted high-temperature lifetimes are shown in \crefformat{figure}{Fig.~#2#1{(b)}#3}\cref{fig.High-T Results}.  It is not obvious from the three relevant data points whether this high-temperature mechanism induces decay out of the $\mso$ states as it does out of the $\msz$ states.

The previous explanation given for the high-temperature decay is inconsistent with the model outlined in this work because there is no SO coupling between $\Exy$ and $\As$ and because phonons cannot couple $\Exy$ to $\Aone$, which is SO-coupled to $\As$.  Instead, we propose that one of two mechanisms may be responsible. First, this mechanism may be a SO-mediated transition from $\Exy$ to the excited spin-singlet $|^1E_{x,y}\rangle$ states.  $\Exy$ are SO-coupled to $|^1E_{x,y}\rangle$ just as $\Aone$ is SO-coupled to $\As$, so this mechanism would be exactly analogous to the ISC mechanism from $\Aone$.  This mechanism may induce decay from the $\mso$ states because $\Exy$ are coupled to $\Eonetwo$ by a spin-spin interaction \cite{Maze2011,Doherty2011}.  Because $\Ey$ and $\Eonetwo$ exhibit a level anticrossing when the strain-induced $\Ex-\Ey$ splitting is approximately 7 GHz \cite{Robledo2011a}, the spin-spin-mediated decay rate out of the $\mso$ states would be highly sensitive to crystal strain, making it difficult to predict a value of $\epsilon$ for this mechanism. Alternatively, this mechanism may be a direct nonradiative transition to the $\At$ ground state. $E$-symmetric phonons couple the states of the $\Et$ and $\At$ manifolds. Consequently,  the calculation of this rate would be similar to that of the ISC rate from $\Eonetwo$, except that this mechanism would be a first-order process that does not involve SO coupling in addition to phononic coupling.  This mechanism would be spin-conserving and couple to all $\Et$ states equally, implying $\epsilon=1$.

While the theory of these two possible high-temperature decay mechanisms is beyond the scope of this work, we emphasize that they can be distinguished by further high-temperature spin-resolved fluorescence lifetime experiments, which could reduce the uncertainty in the parameter $\epsilon$ and measure the variation of $\epsilon$ with strain.

\section{Conclusion}
\label{sec.Conclusion}

We have presented a microscopic model of the ISC mechanism, which is mediated by spin-orbit coupling and, for some initial states, the emission of an $E$-symmetric phonon.  We have quantitatively shown the model's predictions to be consistent with experimental observations, and have used this comparison to place bounds on the singlet-triplet energy spacing, an important but previously unknown property of the NV center's level structure.


The bounds we place on $\Delta$ have implications for efforts to engineer the ISC rate, which could increase the measurement readout visibility between $|0\rangle$ and $|\pm1\rangle$ by increasing the ISC rate out of the $\mso$ states.  We estimate that the ISC rate increases by $2\pi\times\left(0.15\pm0.05\right)$ MHz/meV as $\Delta$ decreases.  Because it is the competition of $\GISC$ with $\GRad=2\pi\times\left(13.2\pm0.5\right)$ MHz that gives rise to the fluorescence contrast between $|0\rangle$ and $|\pm1\rangle$, our finding suggests that a large ($\sim100-200$ meV) reduction in $\Delta$ would be needed to achieve an appreciable improvement in nonresonant readout fidelity.

The application of isotropic hydrostatic pressure can induce large shifts ($\sim400$ meV \cite{Doherty2014}), but the direction of this shift corresponds to an increase in $\Delta$, which would reduce the ISC rate.  Conversely, the application of uniaxial strain (specifically along the $[111]$ crystal axis) can induce energy shifts that reduce $\Delta$, but the structural integrity of bulk diamond under unavoidable shear forces limits these shifts to $\sim10$ meV \cite{Davies1976,Rogers2014}.  It may be possible, however, to induce a suitably large shift in diamond nanofabricated structures \cite{MacQuarrie2013,Ovartchaiyapong2014a},
wherein the application of a small local force may cause a large strain at the NV center.  One could measure this shift in the ISC rate either by the techniques employed in Ref. \onlinecite{Goldman2014} or by measuring the visibility of ground state optically detected magnetic resonance (ODMR) as in Ref. \onlinecite{Doherty2014}.  By identifying methods of strain application that maximize ODMR visibility, such an experiment could significantly enhance the spin initialization and readout techniques upon which room-temperature NV center applications depend.

\section*{Acknowledgments}

The authors would like to thank J. Maze, A. Gali, E. Togan, A. Akimov, D. Sukachev, Q. Unterreithmeier, A. Zibrov, and  A. Palyi for stimulating discussions and experimental contributions. This work was supported by NSF, CUA, the DARPA QUEST program, AFOSR MURI, ARC, Element Six, and the Packard Foundation. A. K. acknowledges support from the Alexander von Humboldt Foundation.

\section*{Appendix}

We now consider the contribution to $\GEonetwo$ due to an ISC process that uses $\Es$ as intermediate states instead of $\Aone$.  In this process, $\Eonetwo$ are SO-coupled to $\Es$, and phonons of $E$ symmetry couple $\Es$ to $\As$.  We modify Eq. \ref{eq.Initial GEonetwo}, with which we begin our calculation of $\GEonetwo$, to include the contribution of this secondary process, finding
\begin{align}
\Gamma_{E_{1,2}} = & \; 2 \pi \hbar^3 \sum\limits_{n,p,k} \left|\sqrt{2} \, \lambda_\perp \lambda_{p,k}\right|^2 \: \delta\hspace{-1 pt}\left(\nu_n + \omega_k - \Delta\right) \nonumber \\
& \times \Bigg| \frac{\langle \chi_0 | \chi^\prime_{\nu_n} \rangle}{\omega_k}
- \sum\limits_{n^\prime} \frac{2 \langle \chi_0 | \chi^{\prime\prime}_{\nu_{n^\prime}} \rangle \langle \chi^{\prime\prime}_{\nu_{n^\prime}} | \chi^\prime_{\nu_n} \rangle}
{\Delta+\Delta^\prime-\nu_{n^\prime}} \Bigg|^2 \hspace{-2 pt},
\end{align}
where $| \chi^{\prime\prime}_{\nu_{n^\prime}} \rangle$ is an excited vibrational level of $\Es$ that is separated from $\Es$ by an energy spacing $\nu_{n^\prime}$ and $\Delta^\prime$ is the $\As-\Es$ energy splitting.  We add the two contributions' amplitudes instead of their magnitudes because the final states arrived at by both processes are identical for given values of $n$, $p$, and $k$.  The two terms have opposite signs because the phonon is emitted first (second) for the mechanism using $\Aone$ ($\Eonetwo$) as an intermediate state, making the detuning denominator $\mathcal{E}_i-\hat{H}_0$ negative (positive).

Because the energy spacing $\Delta+\Delta^\prime$ ($387\pm43+1190$ meV \cite{Acosta2010a}) is large compared to the extent of the phonon sideband ($\sim500$ meV \cite{Davies1974,Kehayias2013}), we make the simplifying assumption that $\langle \chi_0 | \chi^{\prime\prime}_{\nu_{n^\prime}} \rangle$ is only appreciable for $\nu_{n^\prime}\ll\Delta+\Delta^\prime$.  We use this approximation, the expression for the phonon spectral density in the acoustic regime given by Eq. \ref{eq.Spectral Density}, and the identity operator $\hat{1} = \sum_{n^\prime} | \chi^{\prime\prime}_{\nu_{n^\prime}} \rangle \langle \chi^{\prime\prime}_{\nu_{n^\prime}} |$ to find
\begin{align}
\Gamma_{E_{1,2}} = & \; 8 \: \hbar^2 \lambda_\perp^2 \eta \int\limits_0^\Omega \: \omega^3 \: \sum\limits_{n} \delta\hspace{-1 pt}\left(\nu_n + \omega - \Delta\right) \nonumber \\
& \times \Bigg| \left(\frac{1}{\omega}-\frac{2}{\Delta+\Delta^\prime}\right)\langle \chi_0 | \chi^\prime_{\nu_n} \rangle \nonumber \\
& \;\;\;\; - \sum\limits_{n^\prime} \frac{2 \nu_{n^\prime} \langle \chi_0 | \chi^{\prime\prime}_{\nu_{n^\prime}} \rangle \langle \chi^{\prime\prime}_{\nu_{n^\prime}} | \chi^\prime_{\nu_n} \rangle}{(\Delta+\Delta^\prime)^2}
\Bigg|^2 \mathrm{d}\omega.
\end{align}

Finally, we neglect the last term, which is a second-order correction in $\nu_{n^\prime},\omega\ll\Delta+\Delta^\prime$, to find
\begin{align}
\Gamma_{E_{1,2}} = 8 \: \hbar^2 \lambda_\perp^2 \eta \int\limits_0^\Omega \: & \omega^3 \: \left(\frac{1}{\omega}-\frac{2}{\Delta+\Delta^\prime}\right)^2 \nonumber \\
& \times F\hspace{-1.5 pt}\left(\Delta-\omega\right) \mathrm{d}\omega.
\end{align}
This expression trivially reduces to Eq. \ref{eq.E12 ISC} for $\Delta^\prime\rightarrow\infty$.  This modification constitutes a $\sim15\%$ downward correction to $\GEonetwo$ for the relevant values of $\Delta$ and $\Omega$, as shown in Fig. \ref{fig.ISC ratio vs data}.


\begin{thebibliography}{46}%
\makeatletter
\providecommand \@ifxundefined [1]{%
 \@ifx{#1\undefined}
}%
\providecommand \@ifnum [1]{%
 \ifnum #1\expandafter \@firstoftwo
 \else \expandafter \@secondoftwo
 \fi
}%
\providecommand \@ifx [1]{%
 \ifx #1\expandafter \@firstoftwo
 \else \expandafter \@secondoftwo
 \fi
}%
\providecommand \natexlab [1]{#1}%
\providecommand \enquote  [1]{``#1''}%
\providecommand \bibnamefont  [1]{#1}%
\providecommand \bibfnamefont [1]{#1}%
\providecommand \citenamefont [1]{#1}%
\providecommand \href@noop [0]{\@secondoftwo}%
\providecommand \href [0]{\begingroup \@sanitize@url \@href}%
\providecommand \@href[1]{\@@startlink{#1}\@@href}%
\providecommand \@@href[1]{\endgroup#1\@@endlink}%
\providecommand \@sanitize@url [0]{\catcode `\\12\catcode `\$12\catcode
  `\&12\catcode `\#12\catcode `\^12\catcode `\_12\catcode `\%12\relax}%
\providecommand \@@startlink[1]{}%
\providecommand \@@endlink[0]{}%
\providecommand \url  [0]{\begingroup\@sanitize@url \@url }%
\providecommand \@url [1]{\endgroup\@href {#1}{\urlprefix }}%
\providecommand \urlprefix  [0]{URL }%
\providecommand \Eprint [0]{\href }%
\providecommand \doibase [0]{http://dx.doi.org/}%
\providecommand \selectlanguage [0]{\@gobble}%
\providecommand \bibinfo  [0]{\@secondoftwo}%
\providecommand \bibfield  [0]{\@secondoftwo}%
\providecommand \translation [1]{[#1]}%
\providecommand \BibitemOpen [0]{}%
\providecommand \bibitemStop [0]{}%
\providecommand \bibitemNoStop [0]{.\EOS\space}%
\providecommand \EOS [0]{\spacefactor3000\relax}%
\providecommand \BibitemShut  [1]{\csname bibitem#1\endcsname}%
\let\auto@bib@innerbib\@empty
\bibitem [{\citenamefont {Kucsko}\ \emph {et~al.}(2013)\citenamefont {Kucsko},
  \citenamefont {Maurer}, \citenamefont {Yao}, \citenamefont {Kubo},
  \citenamefont {Noh}, \citenamefont {Lo}, \citenamefont {Park},\ and\
  \citenamefont {Lukin}}]{Kucsko2013}%
  \BibitemOpen
  \bibfield  {author} {\bibinfo {author} {\bibfnamefont {G.}~\bibnamefont
  {Kucsko}}, \bibinfo {author} {\bibfnamefont {P.~C.}\ \bibnamefont {Maurer}},
  \bibinfo {author} {\bibfnamefont {N.~Y.}\ \bibnamefont {Yao}}, \bibinfo
  {author} {\bibfnamefont {M.}~\bibnamefont {Kubo}}, \bibinfo {author}
  {\bibfnamefont {H.~J.}\ \bibnamefont {Noh}}, \bibinfo {author} {\bibfnamefont
  {P.~K.}\ \bibnamefont {Lo}}, \bibinfo {author} {\bibfnamefont
  {H.}~\bibnamefont {Park}}, \ and\ \bibinfo {author} {\bibfnamefont {M.~D.}\
  \bibnamefont {Lukin}},\ }\href {\doibase 10.1038/nature12373} {\bibfield
  {journal} {\bibinfo  {journal} {Nature (London)}\ }\textbf {\bibinfo {volume}
  {500}},\ \bibinfo {pages} {54} (\bibinfo {year} {2013})}\BibitemShut
  {NoStop}%
\bibitem [{\citenamefont {{Le Sage}}\ \emph {et~al.}(2013)\citenamefont {{Le
  Sage}}, \citenamefont {Arai}, \citenamefont {Glenn}, \citenamefont
  {DeVience}, \citenamefont {Pham}, \citenamefont {Rahn-Lee}, \citenamefont
  {Lukin}, \citenamefont {Yacoby}, \citenamefont {Komeili},\ and\ \citenamefont
  {Walsworth}}]{LeSage2013}%
  \BibitemOpen
  \bibfield  {author} {\bibinfo {author} {\bibfnamefont {D.}~\bibnamefont {{Le
  Sage}}}, \bibinfo {author} {\bibfnamefont {K.}~\bibnamefont {Arai}}, \bibinfo
  {author} {\bibfnamefont {D.~R.}\ \bibnamefont {Glenn}}, \bibinfo {author}
  {\bibfnamefont {S.~J.}\ \bibnamefont {DeVience}}, \bibinfo {author}
  {\bibfnamefont {L.~M.}\ \bibnamefont {Pham}}, \bibinfo {author}
  {\bibfnamefont {L.}~\bibnamefont {Rahn-Lee}}, \bibinfo {author}
  {\bibfnamefont {M.~D.}\ \bibnamefont {Lukin}}, \bibinfo {author}
  {\bibfnamefont {A.}~\bibnamefont {Yacoby}}, \bibinfo {author} {\bibfnamefont
  {A.}~\bibnamefont {Komeili}}, \ and\ \bibinfo {author} {\bibfnamefont
  {R.~L.}\ \bibnamefont {Walsworth}},\ }\href {\doibase 10.1038/nature12072}
  {\bibfield  {journal} {\bibinfo  {journal} {Nature (London)}\ }\textbf
  {\bibinfo {volume} {496}},\ \bibinfo {pages} {486} (\bibinfo {year}
  {2013})}\BibitemShut {NoStop}%
\bibitem [{\citenamefont {Grinolds}\ \emph {et~al.}(2013)\citenamefont
  {Grinolds}, \citenamefont {Hong}, \citenamefont {Maletinsky}, \citenamefont
  {Luan}, \citenamefont {Lukin}, \citenamefont {Walsworth},\ and\ \citenamefont
  {Yacoby}}]{Grinolds2013}%
  \BibitemOpen
  \bibfield  {author} {\bibinfo {author} {\bibfnamefont {M.~S.}\ \bibnamefont
  {Grinolds}}, \bibinfo {author} {\bibfnamefont {S.}~\bibnamefont {Hong}},
  \bibinfo {author} {\bibfnamefont {P.}~\bibnamefont {Maletinsky}}, \bibinfo
  {author} {\bibfnamefont {L.}~\bibnamefont {Luan}}, \bibinfo {author}
  {\bibfnamefont {M.~D.}\ \bibnamefont {Lukin}}, \bibinfo {author}
  {\bibfnamefont {R.~L.}\ \bibnamefont {Walsworth}}, \ and\ \bibinfo {author}
  {\bibfnamefont {A.}~\bibnamefont {Yacoby}},\ }\href {\doibase
  10.1038/nphys2543} {\bibfield  {journal} {\bibinfo  {journal} {Nature Phys.}\
  }\textbf {\bibinfo {volume} {9}},\ \bibinfo {pages} {215} (\bibinfo {year}
  {2013})}\BibitemShut {NoStop}%
\bibitem [{\citenamefont {Ermakova}\ \emph {et~al.}(2013)\citenamefont
  {Ermakova}, \citenamefont {Pramanik}, \citenamefont {Cai}, \citenamefont
  {Algara-Siller}, \citenamefont {Kaiser}, \citenamefont {Weil}, \citenamefont
  {Tzeng}, \citenamefont {Chang}, \citenamefont {McGuinness}, \citenamefont
  {Plenio}, \citenamefont {Naydenov},\ and\ \citenamefont
  {Jelezko}}]{Ermakova2013}%
  \BibitemOpen
  \bibfield  {author} {\bibinfo {author} {\bibfnamefont {A.}~\bibnamefont
  {Ermakova}}, \bibinfo {author} {\bibfnamefont {G.}~\bibnamefont {Pramanik}},
  \bibinfo {author} {\bibfnamefont {J.-M.}\ \bibnamefont {Cai}}, \bibinfo
  {author} {\bibfnamefont {G.}~\bibnamefont {Algara-Siller}}, \bibinfo {author}
  {\bibfnamefont {U.}~\bibnamefont {Kaiser}}, \bibinfo {author} {\bibfnamefont
  {T.}~\bibnamefont {Weil}}, \bibinfo {author} {\bibfnamefont {Y.-K.}\
  \bibnamefont {Tzeng}}, \bibinfo {author} {\bibfnamefont {H.~C.}\ \bibnamefont
  {Chang}}, \bibinfo {author} {\bibfnamefont {L.~P.}\ \bibnamefont
  {McGuinness}}, \bibinfo {author} {\bibfnamefont {M.~B.}\ \bibnamefont
  {Plenio}}, \bibinfo {author} {\bibfnamefont {B.}~\bibnamefont {Naydenov}}, \
  and\ \bibinfo {author} {\bibfnamefont {F.}~\bibnamefont {Jelezko}},\ }\href
  {\doibase 10.1021/nl4015233} {\bibfield  {journal} {\bibinfo  {journal} {Nano
  Lett.}\ }\textbf {\bibinfo {volume} {13}},\ \bibinfo {pages} {3305} (\bibinfo
  {year} {2013})}\BibitemShut {NoStop}%
\bibitem [{\citenamefont {Sushkov}\ \emph {et~al.}()\citenamefont {Sushkov},
  \citenamefont {Chisholm}, \citenamefont {Lovchinsky}, \citenamefont {Kubo},
  \citenamefont {Lo}, \citenamefont {Bennett}, \citenamefont {Hunger},
  \citenamefont {Akimov}, \citenamefont {Walsworth}, \citenamefont {Park},\
  and\ \citenamefont {Lukin}}]{Sushkov2013}%
  \BibitemOpen
  \bibfield  {author} {\bibinfo {author} {\bibfnamefont {A.~O.}\ \bibnamefont
  {Sushkov}}, \bibinfo {author} {\bibfnamefont {N.}~\bibnamefont {Chisholm}},
  \bibinfo {author} {\bibfnamefont {I.}~\bibnamefont {Lovchinsky}}, \bibinfo
  {author} {\bibfnamefont {M.}~\bibnamefont {Kubo}}, \bibinfo {author}
  {\bibfnamefont {P.~K.}\ \bibnamefont {Lo}}, \bibinfo {author} {\bibfnamefont
  {S.~D.}\ \bibnamefont {Bennett}}, \bibinfo {author} {\bibfnamefont
  {D.}~\bibnamefont {Hunger}}, \bibinfo {author} {\bibfnamefont
  {A.}~\bibnamefont {Akimov}}, \bibinfo {author} {\bibfnamefont {R.~L.}\
  \bibnamefont {Walsworth}}, \bibinfo {author} {\bibfnamefont {H.}~\bibnamefont
  {Park}}, \ and\ \bibinfo {author} {\bibfnamefont {M.~D.}\ \bibnamefont
  {Lukin}},\ }\href@noop {} {\ }\Eprint {http://arxiv.org/abs/1311.1801}
  {arXiv:1311.1801} \BibitemShut {NoStop}%
\bibitem [{\citenamefont {Dolde}\ \emph {et~al.}(2011)\citenamefont {Dolde},
  \citenamefont {Fedder}, \citenamefont {Doherty}, \citenamefont {N\"{o}bauer},
  \citenamefont {Rempp}, \citenamefont {Balasubramanian}, \citenamefont {Wolf},
  \citenamefont {Reinhard}, \citenamefont {Hollenberg}, \citenamefont
  {Jelezko},\ and\ \citenamefont {Wrachtrup}}]{Dolde2011}%
  \BibitemOpen
  \bibfield  {author} {\bibinfo {author} {\bibfnamefont {F.}~\bibnamefont
  {Dolde}}, \bibinfo {author} {\bibfnamefont {H.}~\bibnamefont {Fedder}},
  \bibinfo {author} {\bibfnamefont {M.~W.}\ \bibnamefont {Doherty}}, \bibinfo
  {author} {\bibfnamefont {T.}~\bibnamefont {N\"{o}bauer}}, \bibinfo {author}
  {\bibfnamefont {F.}~\bibnamefont {Rempp}}, \bibinfo {author} {\bibfnamefont
  {G.}~\bibnamefont {Balasubramanian}}, \bibinfo {author} {\bibfnamefont
  {T.}~\bibnamefont {Wolf}}, \bibinfo {author} {\bibfnamefont {F.}~\bibnamefont
  {Reinhard}}, \bibinfo {author} {\bibfnamefont {L.~C.~L.}\ \bibnamefont
  {Hollenberg}}, \bibinfo {author} {\bibfnamefont {F.}~\bibnamefont {Jelezko}},
  \ and\ \bibinfo {author} {\bibfnamefont {J.}~\bibnamefont {Wrachtrup}},\
  }\href {\doibase 10.1038/nphys1969} {\bibfield  {journal} {\bibinfo
  {journal} {Nature Phys.}\ }\textbf {\bibinfo {volume} {7}},\ \bibinfo {pages}
  {459} (\bibinfo {year} {2011})}\BibitemShut {NoStop}%
\bibitem [{\citenamefont {Doherty}\ \emph {et~al.}(2014)\citenamefont
  {Doherty}, \citenamefont {Struzhkin}, \citenamefont {Simpson}, \citenamefont
  {McGuinness}, \citenamefont {Meng}, \citenamefont {Stacey}, \citenamefont
  {Karle}, \citenamefont {Hemley}, \citenamefont {Manson}, \citenamefont
  {Hollenberg},\ and\ \citenamefont {Prawer}}]{Doherty2014}%
  \BibitemOpen
  \bibfield  {author} {\bibinfo {author} {\bibfnamefont {M.~W.}\ \bibnamefont
  {Doherty}}, \bibinfo {author} {\bibfnamefont {V.~V.}\ \bibnamefont
  {Struzhkin}}, \bibinfo {author} {\bibfnamefont {D.~A.}\ \bibnamefont
  {Simpson}}, \bibinfo {author} {\bibfnamefont {L.~P.}\ \bibnamefont
  {McGuinness}}, \bibinfo {author} {\bibfnamefont {Y.}~\bibnamefont {Meng}},
  \bibinfo {author} {\bibfnamefont {A.}~\bibnamefont {Stacey}}, \bibinfo
  {author} {\bibfnamefont {T.~J.}\ \bibnamefont {Karle}}, \bibinfo {author}
  {\bibfnamefont {R.~J.}\ \bibnamefont {Hemley}}, \bibinfo {author}
  {\bibfnamefont {N.~B.}\ \bibnamefont {Manson}}, \bibinfo {author}
  {\bibfnamefont {L.~C.~L.}\ \bibnamefont {Hollenberg}}, \ and\ \bibinfo
  {author} {\bibfnamefont {S.}~\bibnamefont {Prawer}},\ }\href {\doibase
  10.1103/PhysRevLett.112.047601} {\bibfield  {journal} {\bibinfo  {journal}
  {Phys. Rev. Lett.}\ }\textbf {\bibinfo {volume} {112}},\ \bibinfo {pages}
  {047601} (\bibinfo {year} {2014})}\BibitemShut {NoStop}%
\bibitem [{\citenamefont {Dutt}\ \emph {et~al.}(2007)\citenamefont {Dutt},
  \citenamefont {Childress}, \citenamefont {Jiang}, \citenamefont {Togan},
  \citenamefont {Maze}, \citenamefont {Jelezko}, \citenamefont {Zibrov},
  \citenamefont {Hemmer},\ and\ \citenamefont {Lukin}}]{Dutt2007}%
  \BibitemOpen
  \bibfield  {author} {\bibinfo {author} {\bibfnamefont {M.~V.~G.}\
  \bibnamefont {Dutt}}, \bibinfo {author} {\bibfnamefont {L.}~\bibnamefont
  {Childress}}, \bibinfo {author} {\bibfnamefont {L.}~\bibnamefont {Jiang}},
  \bibinfo {author} {\bibfnamefont {E.}~\bibnamefont {Togan}}, \bibinfo
  {author} {\bibfnamefont {J.}~\bibnamefont {Maze}}, \bibinfo {author}
  {\bibfnamefont {F.}~\bibnamefont {Jelezko}}, \bibinfo {author} {\bibfnamefont
  {A.~S.}\ \bibnamefont {Zibrov}}, \bibinfo {author} {\bibfnamefont {P.~R.}\
  \bibnamefont {Hemmer}}, \ and\ \bibinfo {author} {\bibfnamefont {M.~D.}\
  \bibnamefont {Lukin}},\ }\href {\doibase 10.1126/science.1139831} {\bibfield
  {journal} {\bibinfo  {journal} {Science}\ }\textbf {\bibinfo {volume}
  {316}},\ \bibinfo {pages} {1312} (\bibinfo {year} {2007})}\BibitemShut
  {NoStop}%
\bibitem [{\citenamefont {van~der Sar}\ \emph {et~al.}(2012)\citenamefont
  {van~der Sar}, \citenamefont {Wang}, \citenamefont {Blok}, \citenamefont
  {Bernien}, \citenamefont {Taminiau}, \citenamefont {Toyli}, \citenamefont
  {Lidar}, \citenamefont {Awschalom}, \citenamefont {Hanson},\ and\
  \citenamefont {Dobrovitski}}]{VanderSar2012}%
  \BibitemOpen
  \bibfield  {author} {\bibinfo {author} {\bibfnamefont {T.}~\bibnamefont
  {van~der Sar}}, \bibinfo {author} {\bibfnamefont {Z.~H.}\ \bibnamefont
  {Wang}}, \bibinfo {author} {\bibfnamefont {M.~S.}\ \bibnamefont {Blok}},
  \bibinfo {author} {\bibfnamefont {H.}~\bibnamefont {Bernien}}, \bibinfo
  {author} {\bibfnamefont {T.~H.}\ \bibnamefont {Taminiau}}, \bibinfo {author}
  {\bibfnamefont {D.~M.}\ \bibnamefont {Toyli}}, \bibinfo {author}
  {\bibfnamefont {D.~A.}\ \bibnamefont {Lidar}}, \bibinfo {author}
  {\bibfnamefont {D.~D.}\ \bibnamefont {Awschalom}}, \bibinfo {author}
  {\bibfnamefont {R.}~\bibnamefont {Hanson}}, \ and\ \bibinfo {author}
  {\bibfnamefont {V.~V.}\ \bibnamefont {Dobrovitski}},\ }\href {\doibase
  10.1038/nature10900} {\bibfield  {journal} {\bibinfo  {journal} {Nature
  (London)}\ }\textbf {\bibinfo {volume} {484}},\ \bibinfo {pages} {82}
  (\bibinfo {year} {2012})}\BibitemShut {NoStop}%
\bibitem [{\citenamefont {Maurer}\ \emph {et~al.}(2012)\citenamefont {Maurer},
  \citenamefont {Kucsko}, \citenamefont {Latta}, \citenamefont {Jiang},
  \citenamefont {Yao}, \citenamefont {Bennett}, \citenamefont {Pastawski},
  \citenamefont {Hunger}, \citenamefont {Chisholm}, \citenamefont {Markham},
  \citenamefont {Twitchen}, \citenamefont {Cirac},\ and\ \citenamefont
  {Lukin}}]{Maurer2012a}%
  \BibitemOpen
  \bibfield  {author} {\bibinfo {author} {\bibfnamefont {P.~C.}\ \bibnamefont
  {Maurer}}, \bibinfo {author} {\bibfnamefont {G.}~\bibnamefont {Kucsko}},
  \bibinfo {author} {\bibfnamefont {C.}~\bibnamefont {Latta}}, \bibinfo
  {author} {\bibfnamefont {L.}~\bibnamefont {Jiang}}, \bibinfo {author}
  {\bibfnamefont {N.~Y.}\ \bibnamefont {Yao}}, \bibinfo {author} {\bibfnamefont
  {S.~D.}\ \bibnamefont {Bennett}}, \bibinfo {author} {\bibfnamefont
  {F.}~\bibnamefont {Pastawski}}, \bibinfo {author} {\bibfnamefont
  {D.}~\bibnamefont {Hunger}}, \bibinfo {author} {\bibfnamefont
  {N.}~\bibnamefont {Chisholm}}, \bibinfo {author} {\bibfnamefont
  {M.}~\bibnamefont {Markham}}, \bibinfo {author} {\bibfnamefont {D.~J.}\
  \bibnamefont {Twitchen}}, \bibinfo {author} {\bibfnamefont {J.~I.}\
  \bibnamefont {Cirac}}, \ and\ \bibinfo {author} {\bibfnamefont {M.~D.}\
  \bibnamefont {Lukin}},\ }\href {\doibase 10.1126/science.1220513} {\bibfield
  {journal} {\bibinfo  {journal} {Science}\ }\textbf {\bibinfo {volume}
  {336}},\ \bibinfo {pages} {1283} (\bibinfo {year} {2012})}\BibitemShut
  {NoStop}%
\bibitem [{\citenamefont {Manson}\ \emph {et~al.}(2006)\citenamefont {Manson},
  \citenamefont {Harrison},\ and\ \citenamefont {Sellars}}]{Manson2006}%
  \BibitemOpen
  \bibfield  {author} {\bibinfo {author} {\bibfnamefont {N.~B.}\ \bibnamefont
  {Manson}}, \bibinfo {author} {\bibfnamefont {J.~P.}\ \bibnamefont
  {Harrison}}, \ and\ \bibinfo {author} {\bibfnamefont {M.~J.}\ \bibnamefont
  {Sellars}},\ }\href {\doibase 10.1103/PhysRevB.74.104303} {\bibfield
  {journal} {\bibinfo  {journal} {Phys. Rev. B}\ }\textbf {\bibinfo {volume}
  {74}},\ \bibinfo {pages} {104303} (\bibinfo {year} {2006})}\BibitemShut
  {NoStop}%
\bibitem [{\citenamefont {Toyli}\ \emph {et~al.}(2012)\citenamefont {Toyli},
  \citenamefont {Christle}, \citenamefont {Alkauskas}, \citenamefont {Buckley},
  \citenamefont {{Van de Walle}},\ and\ \citenamefont {Awschalom}}]{Toyli2012}%
  \BibitemOpen
  \bibfield  {author} {\bibinfo {author} {\bibfnamefont {D.~M.}\ \bibnamefont
  {Toyli}}, \bibinfo {author} {\bibfnamefont {D.~J.}\ \bibnamefont {Christle}},
  \bibinfo {author} {\bibfnamefont {A.}~\bibnamefont {Alkauskas}}, \bibinfo
  {author} {\bibfnamefont {B.~B.}\ \bibnamefont {Buckley}}, \bibinfo {author}
  {\bibfnamefont {C.~G.}\ \bibnamefont {{Van de Walle}}}, \ and\ \bibinfo
  {author} {\bibfnamefont {D.~D.}\ \bibnamefont {Awschalom}},\ }\href {\doibase
  10.1103/PhysRevX.2.031001} {\bibfield  {journal} {\bibinfo  {journal} {Phys.
  Rev. X}\ }\textbf {\bibinfo {volume} {2}},\ \bibinfo {pages} {031001}
  (\bibinfo {year} {2012})}\BibitemShut {NoStop}%
\bibitem [{\citenamefont {Batalov}\ \emph {et~al.}(2008)\citenamefont
  {Batalov}, \citenamefont {Zierl}, \citenamefont {Gaebel}, \citenamefont
  {Neumann}, \citenamefont {Chan}, \citenamefont {Balasubramanian},
  \citenamefont {Hemmer}, \citenamefont {Jelezko},\ and\ \citenamefont
  {Wrachtrup}}]{Batalov2008}%
  \BibitemOpen
  \bibfield  {author} {\bibinfo {author} {\bibfnamefont {A.}~\bibnamefont
  {Batalov}}, \bibinfo {author} {\bibfnamefont {C.}~\bibnamefont {Zierl}},
  \bibinfo {author} {\bibfnamefont {T.}~\bibnamefont {Gaebel}}, \bibinfo
  {author} {\bibfnamefont {P.}~\bibnamefont {Neumann}}, \bibinfo {author}
  {\bibfnamefont {I.-Y.}\ \bibnamefont {Chan}}, \bibinfo {author}
  {\bibfnamefont {G.}~\bibnamefont {Balasubramanian}}, \bibinfo {author}
  {\bibfnamefont {P.~R.}\ \bibnamefont {Hemmer}}, \bibinfo {author}
  {\bibfnamefont {F.}~\bibnamefont {Jelezko}}, \ and\ \bibinfo {author}
  {\bibfnamefont {J.}~\bibnamefont {Wrachtrup}},\ }\href {\doibase
  10.1103/PhysRevLett.100.077401} {\bibfield  {journal} {\bibinfo  {journal}
  {Phys. Rev. Lett.}\ }\textbf {\bibinfo {volume} {100}},\ \bibinfo {pages}
  {077401} (\bibinfo {year} {2008})}\BibitemShut {NoStop}%
\bibitem [{\citenamefont {Robledo}\ \emph
  {et~al.}(2011{\natexlab{a}})\citenamefont {Robledo}, \citenamefont {Bernien},
  \citenamefont {van~der Sar},\ and\ \citenamefont {Hanson}}]{Robledo2011}%
  \BibitemOpen
  \bibfield  {author} {\bibinfo {author} {\bibfnamefont {L.}~\bibnamefont
  {Robledo}}, \bibinfo {author} {\bibfnamefont {H.}~\bibnamefont {Bernien}},
  \bibinfo {author} {\bibfnamefont {T.}~\bibnamefont {van~der Sar}}, \ and\
  \bibinfo {author} {\bibfnamefont {R.}~\bibnamefont {Hanson}},\ }\href
  {\doibase 10.1088/1367-2630/13/2/025013} {\bibfield  {journal} {\bibinfo
  {journal} {New J. Phys.}\ }\textbf {\bibinfo {volume} {13}},\ \bibinfo
  {pages} {025013} (\bibinfo {year} {2011}{\natexlab{a}})}\BibitemShut
  {NoStop}%
\bibitem [{\citenamefont {Tetienne}\ \emph {et~al.}(2012)\citenamefont
  {Tetienne}, \citenamefont {Rondin}, \citenamefont {Spinicelli}, \citenamefont
  {Chipaux}, \citenamefont {Debuisschert}, \citenamefont {Roch},\ and\
  \citenamefont {Jacques}}]{Tetienne2012}%
  \BibitemOpen
  \bibfield  {author} {\bibinfo {author} {\bibfnamefont {J.-P.}\ \bibnamefont
  {Tetienne}}, \bibinfo {author} {\bibfnamefont {L.}~\bibnamefont {Rondin}},
  \bibinfo {author} {\bibfnamefont {P.}~\bibnamefont {Spinicelli}}, \bibinfo
  {author} {\bibfnamefont {M.}~\bibnamefont {Chipaux}}, \bibinfo {author}
  {\bibfnamefont {T.}~\bibnamefont {Debuisschert}}, \bibinfo {author}
  {\bibfnamefont {J.-F.}\ \bibnamefont {Roch}}, \ and\ \bibinfo {author}
  {\bibfnamefont {V.}~\bibnamefont {Jacques}},\ }\href {\doibase
  10.1088/1367-2630/14/10/103033} {\bibfield  {journal} {\bibinfo  {journal}
  {New J. Phys.}\ }\textbf {\bibinfo {volume} {14}},\ \bibinfo {pages} {103033}
  (\bibinfo {year} {2012})}\BibitemShut {NoStop}%
\bibitem [{\citenamefont {Goldman}\ \emph {et~al.}()\citenamefont {Goldman},
  \citenamefont {Sipahigil}, \citenamefont {Doherty}, \citenamefont {Yao},
  \citenamefont {Bennett}, \citenamefont {Markham}, \citenamefont {Twitchen},
  \citenamefont {Manson}, \citenamefont {Kubanek},\ and\ \citenamefont
  {Lukin}}]{Goldman2014}%
  \BibitemOpen
  \bibfield  {author} {\bibinfo {author} {\bibfnamefont {M.~L.}\ \bibnamefont
  {Goldman}}, \bibinfo {author} {\bibfnamefont {A.}~\bibnamefont {Sipahigil}},
  \bibinfo {author} {\bibfnamefont {M.~W.}\ \bibnamefont {Doherty}}, \bibinfo
  {author} {\bibfnamefont {N.~Y.}\ \bibnamefont {Yao}}, \bibinfo {author}
  {\bibfnamefont {S.~D.}\ \bibnamefont {Bennett}}, \bibinfo {author}
  {\bibfnamefont {M.}~\bibnamefont {Markham}}, \bibinfo {author} {\bibfnamefont
  {D.~J.}\ \bibnamefont {Twitchen}}, \bibinfo {author} {\bibfnamefont {N.~B.}\
  \bibnamefont {Manson}}, \bibinfo {author} {\bibfnamefont {A.}~\bibnamefont
  {Kubanek}}, \ and\ \bibinfo {author} {\bibfnamefont {M.~D.}\ \bibnamefont
  {Lukin}},\ }\href@noop {} {\ }\Eprint {http://arxiv.org/abs/1406.4065}
  {arXiv:1406.4065} \BibitemShut {NoStop}%
\bibitem [{\citenamefont {Kehayias}\ \emph {et~al.}(2013)\citenamefont
  {Kehayias}, \citenamefont {Doherty}, \citenamefont {English}, \citenamefont
  {Fischer}, \citenamefont {Jarmola}, \citenamefont {Jensen}, \citenamefont
  {Leefer}, \citenamefont {Hemmer}, \citenamefont {Manson},\ and\ \citenamefont
  {Budker}}]{Kehayias2013}%
  \BibitemOpen
  \bibfield  {author} {\bibinfo {author} {\bibfnamefont {P.}~\bibnamefont
  {Kehayias}}, \bibinfo {author} {\bibfnamefont {M.~W.}\ \bibnamefont
  {Doherty}}, \bibinfo {author} {\bibfnamefont {D.}~\bibnamefont {English}},
  \bibinfo {author} {\bibfnamefont {R.}~\bibnamefont {Fischer}}, \bibinfo
  {author} {\bibfnamefont {A.}~\bibnamefont {Jarmola}}, \bibinfo {author}
  {\bibfnamefont {K.}~\bibnamefont {Jensen}}, \bibinfo {author} {\bibfnamefont
  {N.}~\bibnamefont {Leefer}}, \bibinfo {author} {\bibfnamefont
  {P.}~\bibnamefont {Hemmer}}, \bibinfo {author} {\bibfnamefont {N.~B.}\
  \bibnamefont {Manson}}, \ and\ \bibinfo {author} {\bibfnamefont
  {D.}~\bibnamefont {Budker}},\ }\href {\doibase 10.1103/PhysRevB.88.165202}
  {\bibfield  {journal} {\bibinfo  {journal} {Phys. Rev. B}\ }\textbf {\bibinfo
  {volume} {88}},\ \bibinfo {pages} {165202} (\bibinfo {year}
  {2013})}\BibitemShut {NoStop}%
\bibitem [{\citenamefont {Maze}\ \emph {et~al.}(2011)\citenamefont {Maze},
  \citenamefont {Gali}, \citenamefont {Togan}, \citenamefont {Chu},
  \citenamefont {Trifonov}, \citenamefont {Kaxiras},\ and\ \citenamefont
  {Lukin}}]{Maze2011}%
  \BibitemOpen
  \bibfield  {author} {\bibinfo {author} {\bibfnamefont {J.~R.}\ \bibnamefont
  {Maze}}, \bibinfo {author} {\bibfnamefont {A.}~\bibnamefont {Gali}}, \bibinfo
  {author} {\bibfnamefont {E.}~\bibnamefont {Togan}}, \bibinfo {author}
  {\bibfnamefont {Y.}~\bibnamefont {Chu}}, \bibinfo {author} {\bibfnamefont
  {A.}~\bibnamefont {Trifonov}}, \bibinfo {author} {\bibfnamefont
  {E.}~\bibnamefont {Kaxiras}}, \ and\ \bibinfo {author} {\bibfnamefont
  {M.~D.}\ \bibnamefont {Lukin}},\ }\href {\doibase
  10.1088/1367-2630/13/2/025025} {\bibfield  {journal} {\bibinfo  {journal}
  {New J. Phys.}\ }\textbf {\bibinfo {volume} {13}},\ \bibinfo {pages} {025025}
  (\bibinfo {year} {2011})}\BibitemShut {NoStop}%
\bibitem [{\citenamefont {Doherty}\ \emph {et~al.}(2011)\citenamefont
  {Doherty}, \citenamefont {Manson}, \citenamefont {Delaney},\ and\
  \citenamefont {Hollenberg}}]{Doherty2011}%
  \BibitemOpen
  \bibfield  {author} {\bibinfo {author} {\bibfnamefont {M.~W.}\ \bibnamefont
  {Doherty}}, \bibinfo {author} {\bibfnamefont {N.~B.}\ \bibnamefont {Manson}},
  \bibinfo {author} {\bibfnamefont {P.}~\bibnamefont {Delaney}}, \ and\
  \bibinfo {author} {\bibfnamefont {L.~C.~L.}\ \bibnamefont {Hollenberg}},\
  }\href {\doibase 10.1088/1367-2630/13/2/025019} {\bibfield  {journal}
  {\bibinfo  {journal} {New J. Phys.}\ }\textbf {\bibinfo {volume} {13}},\
  \bibinfo {pages} {025019} (\bibinfo {year} {2011})}\BibitemShut {NoStop}%
\bibitem [{\citenamefont {Rogers}\ \emph {et~al.}(2008)\citenamefont {Rogers},
  \citenamefont {Armstrong}, \citenamefont {Sellars},\ and\ \citenamefont
  {Manson}}]{Rogers2008}%
  \BibitemOpen
  \bibfield  {author} {\bibinfo {author} {\bibfnamefont {L.~J.}\ \bibnamefont
  {Rogers}}, \bibinfo {author} {\bibfnamefont {S.}~\bibnamefont {Armstrong}},
  \bibinfo {author} {\bibfnamefont {M.~J.}\ \bibnamefont {Sellars}}, \ and\
  \bibinfo {author} {\bibfnamefont {N.~B.}\ \bibnamefont {Manson}},\ }\href
  {\doibase 10.1088/1367-2630/10/10/103024} {\bibfield  {journal} {\bibinfo
  {journal} {New J. Phys.}\ }\textbf {\bibinfo {volume} {10}},\ \bibinfo
  {pages} {103024} (\bibinfo {year} {2008})}\BibitemShut {NoStop}%
\bibitem [{\citenamefont {Doherty}\ \emph {et~al.}(2013)\citenamefont
  {Doherty}, \citenamefont {Manson}, \citenamefont {Delaney}, \citenamefont
  {Jelezko}, \citenamefont {Wrachtrup},\ and\ \citenamefont
  {Hollenberg}}]{Doherty2013a}%
  \BibitemOpen
  \bibfield  {author} {\bibinfo {author} {\bibfnamefont {M.~W.}\ \bibnamefont
  {Doherty}}, \bibinfo {author} {\bibfnamefont {N.~B.}\ \bibnamefont {Manson}},
  \bibinfo {author} {\bibfnamefont {P.}~\bibnamefont {Delaney}}, \bibinfo
  {author} {\bibfnamefont {F.}~\bibnamefont {Jelezko}}, \bibinfo {author}
  {\bibfnamefont {J.}~\bibnamefont {Wrachtrup}}, \ and\ \bibinfo {author}
  {\bibfnamefont {L.~C.~L.}\ \bibnamefont {Hollenberg}},\ }\href {\doibase
  10.1016/j.physrep.2013.02.001} {\bibfield  {journal} {\bibinfo  {journal}
  {Phys. Rep.}\ }\textbf {\bibinfo {volume} {528}},\ \bibinfo {pages} {1}
  (\bibinfo {year} {2013})}\BibitemShut {NoStop}%
\bibitem [{\citenamefont {Huxter}\ \emph {et~al.}(2013)\citenamefont {Huxter},
  \citenamefont {Oliver}, \citenamefont {Budker},\ and\ \citenamefont
  {Fleming}}]{Huxter2013}%
  \BibitemOpen
  \bibfield  {author} {\bibinfo {author} {\bibfnamefont {V.~M.}\ \bibnamefont
  {Huxter}}, \bibinfo {author} {\bibfnamefont {T.~A.~A.}\ \bibnamefont
  {Oliver}}, \bibinfo {author} {\bibfnamefont {D.}~\bibnamefont {Budker}}, \
  and\ \bibinfo {author} {\bibfnamefont {G.~R.}\ \bibnamefont {Fleming}},\
  }\href {\doibase 10.1038/nphys2753} {\bibfield  {journal} {\bibinfo
  {journal} {Nature Phys.}\ }\textbf {\bibinfo {volume} {9}},\ \bibinfo {pages}
  {744} (\bibinfo {year} {2013})}\BibitemShut {NoStop}%
\bibitem [{\citenamefont {Gali}\ \emph {et~al.}(2011)\citenamefont {Gali},
  \citenamefont {Simon},\ and\ \citenamefont {Lowther}}]{Gali2011}%
  \BibitemOpen
  \bibfield  {author} {\bibinfo {author} {\bibfnamefont {A.}~\bibnamefont
  {Gali}}, \bibinfo {author} {\bibfnamefont {T.}~\bibnamefont {Simon}}, \ and\
  \bibinfo {author} {\bibfnamefont {J.~E.}\ \bibnamefont {Lowther}},\ }\href
  {\doibase 10.1088/1367-2630/13/2/025016} {\bibfield  {journal} {\bibinfo
  {journal} {New J. Phys.}\ }\textbf {\bibinfo {volume} {13}},\ \bibinfo
  {pages} {025016} (\bibinfo {year} {2011})}\BibitemShut {NoStop}%
\bibitem [{\citenamefont {Zhang}\ \emph {et~al.}(2011)\citenamefont {Zhang},
  \citenamefont {Wang}, \citenamefont {Zhu},\ and\ \citenamefont
  {Dobrovitski}}]{Zhang2011}%
  \BibitemOpen
  \bibfield  {author} {\bibinfo {author} {\bibfnamefont {J.}~\bibnamefont
  {Zhang}}, \bibinfo {author} {\bibfnamefont {C.-Z.}\ \bibnamefont {Wang}},
  \bibinfo {author} {\bibfnamefont {Z.~Z.}\ \bibnamefont {Zhu}}, \ and\
  \bibinfo {author} {\bibfnamefont {V.~V.}\ \bibnamefont {Dobrovitski}},\
  }\href {\doibase 10.1103/PhysRevB.84.035211} {\bibfield  {journal} {\bibinfo
  {journal} {Phys. Rev. B}\ }\textbf {\bibinfo {volume} {84}},\ \bibinfo
  {pages} {035211} (\bibinfo {year} {2011})}\BibitemShut {NoStop}%
\bibitem [{\citenamefont {Bransden}\ and\ \citenamefont
  {Joachain}(2003)}]{Bransden2003}%
  \BibitemOpen
  \bibfield  {author} {\bibinfo {author} {\bibfnamefont {B.~H.}\ \bibnamefont
  {Bransden}}\ and\ \bibinfo {author} {\bibfnamefont {C.~J.}\ \bibnamefont
  {Joachain}},\ }\href@noop {} {\emph {\bibinfo {title} {{Physics of Atoms and
  Molecules}}}},\ \bibinfo {edition} {2nd}\ ed.\ (\bibinfo  {publisher}
  {Prentice Hall},\ \bibinfo {address} {Harlow},\ \bibinfo {year}
  {2003})\BibitemShut {NoStop}%
\bibitem [{\citenamefont {Stoneham}(2001)}]{Stoneham2001}%
  \BibitemOpen
  \bibfield  {author} {\bibinfo {author} {\bibfnamefont {A.~M.}\ \bibnamefont
  {Stoneham}},\ }\href {\doibase 10.1093/acprof:oso/9780198507802.001.0001}
  {\emph {\bibinfo {title} {{Theory of Defects in Solids}}}}\ (\bibinfo
  {publisher} {Oxford University Press},\ \bibinfo {address} {Oxford},\
  \bibinfo {year} {2001})\BibitemShut {NoStop}%
\bibitem [{Note1()}]{Note1}%
  \BibitemOpen
  \bibinfo {note} {In group theoretical terms, the polarizations $p=\{1,2\}$
  correspond to the first and second rows of the $E$ irreducible
  representation. Geometrically, phonons of these polarizations induce strain
  of $E_{1,2}^a$ symmetry, as defined in Ref. \protect \rev@citealpnum
  {Maze2011}, which distorts the lattice in directions that are perpendicular
  to the N-V axis.}\BibitemShut {Stop}%
\bibitem [{\citenamefont {Acosta}\ \emph {et~al.}(2010)\citenamefont {Acosta},
  \citenamefont {Jarmola}, \citenamefont {Bauch},\ and\ \citenamefont
  {Budker}}]{Acosta2010a}%
  \BibitemOpen
  \bibfield  {author} {\bibinfo {author} {\bibfnamefont {V.~M.}\ \bibnamefont
  {Acosta}}, \bibinfo {author} {\bibfnamefont {A.}~\bibnamefont {Jarmola}},
  \bibinfo {author} {\bibfnamefont {E.}~\bibnamefont {Bauch}}, \ and\ \bibinfo
  {author} {\bibfnamefont {D.}~\bibnamefont {Budker}},\ }\href {\doibase
  10.1103/PhysRevB.82.201202} {\bibfield  {journal} {\bibinfo  {journal} {Phys.
  Rev. B}\ }\textbf {\bibinfo {volume} {82}},\ \bibinfo {pages} {201202}
  (\bibinfo {year} {2010})}\BibitemShut {NoStop}%
\bibitem [{\citenamefont {Davies}(1974)}]{Davies1974}%
  \BibitemOpen
  \bibfield  {author} {\bibinfo {author} {\bibfnamefont {G.}~\bibnamefont
  {Davies}},\ }\href {\doibase 10.1088/0022-3719/7/20/019} {\bibfield
  {journal} {\bibinfo  {journal} {J. Phys. C}\ }\textbf {\bibinfo {volume}
  {7}},\ \bibinfo {pages} {3797} (\bibinfo {year} {1974})}\BibitemShut
  {NoStop}%
\bibitem [{\citenamefont {Weiss}(2008)}]{Weiss2008}%
  \BibitemOpen
  \bibfield  {author} {\bibinfo {author} {\bibfnamefont {U.}~\bibnamefont
  {Weiss}},\ }\href@noop {} {\emph {\bibinfo {title} {{Quantum Dissipative
  Systems}}}},\ \bibinfo {edition} {3rd}\ ed.\ (\bibinfo  {publisher} {World
  Scientific},\ \bibinfo {address} {Hackensack, NJ},\ \bibinfo {year} {2008})\
  pp.\ \bibinfo {pages} {69--74}\BibitemShut {NoStop}%
\bibitem [{\citenamefont {Fu}\ \emph {et~al.}(2009)\citenamefont {Fu},
  \citenamefont {Santori}, \citenamefont {Barclay}, \citenamefont {Rogers},
  \citenamefont {Manson},\ and\ \citenamefont {Beausoleil}}]{Fu2009}%
  \BibitemOpen
  \bibfield  {author} {\bibinfo {author} {\bibfnamefont {K.-M.~C.}\
  \bibnamefont {Fu}}, \bibinfo {author} {\bibfnamefont {C.}~\bibnamefont
  {Santori}}, \bibinfo {author} {\bibfnamefont {P.~E.}\ \bibnamefont
  {Barclay}}, \bibinfo {author} {\bibfnamefont {L.~J.}\ \bibnamefont {Rogers}},
  \bibinfo {author} {\bibfnamefont {N.~B.}\ \bibnamefont {Manson}}, \ and\
  \bibinfo {author} {\bibfnamefont {R.~G.}\ \bibnamefont {Beausoleil}},\ }\href
  {\doibase 10.1103/PhysRevLett.103.256404} {\bibfield  {journal} {\bibinfo
  {journal} {Phys. Rev. Lett.}\ }\textbf {\bibinfo {volume} {103}},\ \bibinfo
  {pages} {256404} (\bibinfo {year} {2009})}\BibitemShut {NoStop}%
\bibitem [{\citenamefont {Bassett}\ \emph {et~al.}(2014)\citenamefont
  {Bassett}, \citenamefont {Heremans}, \citenamefont {Christle}, \citenamefont
  {Yale}, \citenamefont {Burkard}, \citenamefont {Buckley},\ and\ \citenamefont
  {Awschalom}}]{Bassett2014}%
  \BibitemOpen
  \bibfield  {author} {\bibinfo {author} {\bibfnamefont {L.~C.}\ \bibnamefont
  {Bassett}}, \bibinfo {author} {\bibfnamefont {F.~J.}\ \bibnamefont
  {Heremans}}, \bibinfo {author} {\bibfnamefont {D.~J.}\ \bibnamefont
  {Christle}}, \bibinfo {author} {\bibfnamefont {C.~G.}\ \bibnamefont {Yale}},
  \bibinfo {author} {\bibfnamefont {G.}~\bibnamefont {Burkard}}, \bibinfo
  {author} {\bibfnamefont {B.~B.}\ \bibnamefont {Buckley}}, \ and\ \bibinfo
  {author} {\bibfnamefont {D.~D.}\ \bibnamefont {Awschalom}},\ }\href {\doibase
  10.1126/science.1255541} {\bibfield  {journal} {\bibinfo  {journal}
  {Science}\ }\textbf {\bibinfo {volume} {345}},\ \bibinfo {pages} {1333}
  (\bibinfo {year} {2014})}\BibitemShut {NoStop}%
\bibitem [{\citenamefont {Manson}\ \emph {et~al.}()\citenamefont {Manson},
  \citenamefont {Rogers}, \citenamefont {Doherty},\ and\ \citenamefont
  {Hollenberg}}]{Manson2010}%
  \BibitemOpen
  \bibfield  {author} {\bibinfo {author} {\bibfnamefont {N.~B.}\ \bibnamefont
  {Manson}}, \bibinfo {author} {\bibfnamefont {L.~J.}\ \bibnamefont {Rogers}},
  \bibinfo {author} {\bibfnamefont {M.~W.}\ \bibnamefont {Doherty}}, \ and\
  \bibinfo {author} {\bibfnamefont {L.~C.~L.}\ \bibnamefont {Hollenberg}},\
  }\href@noop {} {\ }\Eprint {http://arxiv.org/abs/1011.2840} {arXiv:1011.2840}
  \BibitemShut {NoStop}%
\bibitem [{\citenamefont {Doherty}()}]{Doherty2012}%
  \BibitemOpen
  \bibfield  {author} {\bibinfo {author} {\bibfnamefont {M.~W.}\ \bibnamefont
  {Doherty}},\ }\href@noop {} {\bibinfo {type} {{Ph.D. thesis}}},\ \bibinfo
  {school} {University of Melbourne, 2012}\BibitemShut {NoStop}%
\bibitem [{Del()}]{DeltaRangeFootnote}%
  \BibitemOpen
  \href@noop {} {}\bibinfo {note} {The bounds \unexpanded{$\Delta_-=344$} meV
  and \unexpanded{$\Delta_+=430$} meV shown in Fig. \ref{fig.A1 ISC vs data}
  take into account both the large uncertainty in \unexpanded{$\lambda_\perp$}
  and the relatively small uncertainty in the measured \unexpanded{$\GAone$}
  rate. In Fig. \ref{fig.ISC ratio vs data}, the uncertainty in the measured
  \unexpanded{$\GAone$} rate is already incorporated into the uncertainty in
  the measured \unexpanded{$\GEonetwo/\GAone$} ratio. Therefore, we use the
  slightly narrower bounds 349 meV and 427 meV, which reflect only the
  uncertainty in $\lambda_\perp$.}\BibitemShut {Stop}%
\bibitem [{Ome()}]{OmegaRangeFootnote}%
  \BibitemOpen
  \href@noop {} {}\bibinfo {note} {\unexpanded{$\Omega$} is a non-analytic
  function of \unexpanded{$\Delta$} (the \unexpanded{$\Aone-\As$} energy
  spacing), \unexpanded{$\eta$} (which parameterizes the electron-phonon
  coupling strength), and the measured \unexpanded{$\GEonetwo/\GAone$} ratio.
  We account for the uncertainties in each of these quantities by choosing the
  combinations of \unexpanded{$\eta\pm\delta\eta$},
  \unexpanded{$\Delta\pm\delta\Delta$} (see above footnote), and
  \unexpanded{$\GEonetwo/\GAone\pm\delta\left(\GEonetwo/\GAone\right)$} that
  minimize and maximize \unexpanded{$\Omega$}.}\BibitemShut {Stop}%
\bibitem [{Tri()}]{TripletEPSBFootnote}%
  \BibitemOpen
  \href@noop {} {}\bibinfo {note} {We compare our extracted value of $\Omega$
  to the $\At\rightarrow\Et$ absorption PSB because this sideband spectrum
  reflects the phonon spectral density corresponding to electronic states in
  the $\Et$ manifold. This is in contrast to Sec. \ref{sec.Quant ISC from A1},
  wherein we use the $\Et\rightarrow\At$ emission PSB to extract information
  about the phonon spectral density in $\At$ (and in $\As$, by
  extension).}\BibitemShut {Stop}%
\bibitem [{\citenamefont {Pavone}\ \emph {et~al.}(1993)\citenamefont {Pavone},
  \citenamefont {Karch}, \citenamefont {Sch\"{u}tt}, \citenamefont {Strauch},
  \citenamefont {Windl}, \citenamefont {Giannozzi},\ and\ \citenamefont
  {Baroni}}]{Pavone1993}%
  \BibitemOpen
  \bibfield  {author} {\bibinfo {author} {\bibfnamefont {P.}~\bibnamefont
  {Pavone}}, \bibinfo {author} {\bibfnamefont {K.}~\bibnamefont {Karch}},
  \bibinfo {author} {\bibfnamefont {O.}~\bibnamefont {Sch\"{u}tt}}, \bibinfo
  {author} {\bibfnamefont {D.}~\bibnamefont {Strauch}}, \bibinfo {author}
  {\bibfnamefont {W.}~\bibnamefont {Windl}}, \bibinfo {author} {\bibfnamefont
  {P.}~\bibnamefont {Giannozzi}}, \ and\ \bibinfo {author} {\bibfnamefont
  {S.}~\bibnamefont {Baroni}},\ }\href
  {http://prb.aps.org/abstract/PRB/v48/i5/p3156\_1} {\bibfield  {journal}
  {\bibinfo  {journal} {Phys. Rev. B}\ }\textbf {\bibinfo {volume} {48}},\
  \bibinfo {pages} {3156} (\bibinfo {year} {1993})}\BibitemShut {NoStop}%
\bibitem [{\citenamefont {Warren}\ \emph {et~al.}(1967)\citenamefont {Warren},
  \citenamefont {Yarnell}, \citenamefont {Dolling},\ and\ \citenamefont
  {Cowley}}]{Warren1967}%
  \BibitemOpen
  \bibfield  {author} {\bibinfo {author} {\bibfnamefont {J.~L.}\ \bibnamefont
  {Warren}}, \bibinfo {author} {\bibfnamefont {J.~L.}\ \bibnamefont {Yarnell}},
  \bibinfo {author} {\bibfnamefont {G.}~\bibnamefont {Dolling}}, \ and\
  \bibinfo {author} {\bibfnamefont {R.~A.}\ \bibnamefont {Cowley}},\ }\href
  {http://prola.aps.org/abstract/PR/v158/i3/p805\_1} {\bibfield  {journal}
  {\bibinfo  {journal} {Phys. Rev.}\ }\textbf {\bibinfo {volume} {158}},\
  \bibinfo {pages} {805} (\bibinfo {year} {1967})}\BibitemShut {NoStop}%
\bibitem [{\citenamefont {Rogers}\ \emph {et~al.}(2009)\citenamefont {Rogers},
  \citenamefont {McMurtrie}, \citenamefont {Sellars},\ and\ \citenamefont
  {Manson}}]{Rogers2009}%
  \BibitemOpen
  \bibfield  {author} {\bibinfo {author} {\bibfnamefont {L.~J.}\ \bibnamefont
  {Rogers}}, \bibinfo {author} {\bibfnamefont {R.~L.}\ \bibnamefont
  {McMurtrie}}, \bibinfo {author} {\bibfnamefont {M.~J.}\ \bibnamefont
  {Sellars}}, \ and\ \bibinfo {author} {\bibfnamefont {N.~B.}\ \bibnamefont
  {Manson}},\ }\href {\doibase 10.1088/1367-2630/11/6/063007} {\bibfield
  {journal} {\bibinfo  {journal} {New J. Phys.}\ }\textbf {\bibinfo {volume}
  {11}},\ \bibinfo {pages} {063007} (\bibinfo {year} {2009})}\BibitemShut
  {NoStop}%
\bibitem [{Note2()}]{Note2}%
  \BibitemOpen
  \bibinfo {note} {In terms of the Mott-Seitz formula given in Ref. \protect
  \rev@citealpnum {Toyli2012}, the high-temperature decay rate is given by
  $\Gamma _\protect \mathrm {HT}\left (T\right ) = s\protect \tmspace
  +\thinmuskip {.1667em}\Gamma _{\protect \mathrm {Rad}}\protect \tmspace
  +\thinmuskip {.1667em}e^{-\Delta E/k_BT}$, where $s$ is the frequency factor
  and $\Delta E$ is the energy barrier for the nonradiative process. We find
  $\Delta E = 0.94\pm 0.32$ eV and $s<5.8\times 10^7$, which differ
  significantly from the values given in Ref. \protect \rev@citealpnum
  {Toyli2012}. This disagreement is not unexpected, however, because we assume
  a low-temperature lifetime of 12.0 ns whereas Toyli et al. fit a
  low-temperature lifetime of $13.4\pm 0.6$ ns. Our fit therefore exhibits a
  sharper turn-on, resulting in a higher value of $\Delta E$; the frequency
  factor $s$, which sets the vertical scaling and is extremely sensitive to
  $\Delta E$, is correspondingly much larger.}\BibitemShut {Stop}%
\bibitem [{\citenamefont {Robledo}\ \emph
  {et~al.}(2011{\natexlab{b}})\citenamefont {Robledo}, \citenamefont
  {Childress}, \citenamefont {Bernien}, \citenamefont {Hensen}, \citenamefont
  {Alkemade},\ and\ \citenamefont {Hanson}}]{Robledo2011a}%
  \BibitemOpen
  \bibfield  {author} {\bibinfo {author} {\bibfnamefont {L.}~\bibnamefont
  {Robledo}}, \bibinfo {author} {\bibfnamefont {L.}~\bibnamefont {Childress}},
  \bibinfo {author} {\bibfnamefont {H.}~\bibnamefont {Bernien}}, \bibinfo
  {author} {\bibfnamefont {B.}~\bibnamefont {Hensen}}, \bibinfo {author}
  {\bibfnamefont {P.~F.~A.}\ \bibnamefont {Alkemade}}, \ and\ \bibinfo {author}
  {\bibfnamefont {R.}~\bibnamefont {Hanson}},\ }\href {\doibase
  10.1038/nature10401} {\bibfield  {journal} {\bibinfo  {journal} {Nature
  (London)}\ }\textbf {\bibinfo {volume} {477}},\ \bibinfo {pages} {574}
  (\bibinfo {year} {2011}{\natexlab{b}})}\BibitemShut {NoStop}%
\bibitem [{\citenamefont {Davies}\ and\ \citenamefont
  {Hamer}(1976)}]{Davies1976}%
  \BibitemOpen
  \bibfield  {author} {\bibinfo {author} {\bibfnamefont {G.}~\bibnamefont
  {Davies}}\ and\ \bibinfo {author} {\bibfnamefont {M.~F.}\ \bibnamefont
  {Hamer}},\ }\href {\doibase 10.1098/rspa.1976.0039} {\bibfield  {journal}
  {\bibinfo  {journal} {Proc. R. Soc. A}\ }\textbf {\bibinfo {volume} {348}},\
  \bibinfo {pages} {285} (\bibinfo {year} {1976})}\BibitemShut {NoStop}%
\bibitem [{\citenamefont {Rogers}\ \emph {et~al.}(2014)\citenamefont {Rogers},
  \citenamefont {Doherty}, \citenamefont {Barson}, \citenamefont {Onoda},
  \citenamefont {Ohshima},\ and\ \citenamefont {Manson}}]{Rogers2014}%
  \BibitemOpen
  \bibfield  {author} {\bibinfo {author} {\bibfnamefont {L.~J.}\ \bibnamefont
  {Rogers}}, \bibinfo {author} {\bibfnamefont {M.~W.}\ \bibnamefont {Doherty}},
  \bibinfo {author} {\bibfnamefont {M.~S.~J.}\ \bibnamefont {Barson}}, \bibinfo
  {author} {\bibfnamefont {S.}~\bibnamefont {Onoda}}, \bibinfo {author}
  {\bibfnamefont {T.}~\bibnamefont {Ohshima}}, \ and\ \bibinfo {author}
  {\bibfnamefont {N.~B.}\ \bibnamefont {Manson}},\ }\href {\doibase
  10.1088/1367-2630/17/1/013048} {\bibfield  {journal} {\bibinfo  {journal}
  {New J. Phys.}\ }\textbf {\bibinfo {volume} {17}},\ \bibinfo {pages} {013048}
  (\bibinfo {year} {2014})}\BibitemShut {NoStop}%
\bibitem [{\citenamefont {MacQuarrie}\ \emph {et~al.}(2013)\citenamefont
  {MacQuarrie}, \citenamefont {Gosavi}, \citenamefont {Jungwirth},
  \citenamefont {Bhave},\ and\ \citenamefont {Fuchs}}]{MacQuarrie2013}%
  \BibitemOpen
  \bibfield  {author} {\bibinfo {author} {\bibfnamefont {E.~R.}\ \bibnamefont
  {MacQuarrie}}, \bibinfo {author} {\bibfnamefont {T.~A.}\ \bibnamefont
  {Gosavi}}, \bibinfo {author} {\bibfnamefont {N.~R.}\ \bibnamefont
  {Jungwirth}}, \bibinfo {author} {\bibfnamefont {S.~A.}\ \bibnamefont
  {Bhave}}, \ and\ \bibinfo {author} {\bibfnamefont {G.~D.}\ \bibnamefont
  {Fuchs}},\ }\href {\doibase 10.1103/PhysRevLett.111.227602} {\bibfield
  {journal} {\bibinfo  {journal} {Phys. Rev. Lett.}\ }\textbf {\bibinfo
  {volume} {111}},\ \bibinfo {pages} {227602} (\bibinfo {year}
  {2013})}\BibitemShut {NoStop}%
\bibitem [{\citenamefont {Ovartchaiyapong}\ \emph {et~al.}(2014)\citenamefont
  {Ovartchaiyapong}, \citenamefont {Lee}, \citenamefont {Myers},\ and\
  \citenamefont {{Bleszynski Jayich}}}]{Ovartchaiyapong2014a}%
  \BibitemOpen
  \bibfield  {author} {\bibinfo {author} {\bibfnamefont {P.}~\bibnamefont
  {Ovartchaiyapong}}, \bibinfo {author} {\bibfnamefont {K.~W.}\ \bibnamefont
  {Lee}}, \bibinfo {author} {\bibfnamefont {B.~A.}\ \bibnamefont {Myers}}, \
  and\ \bibinfo {author} {\bibfnamefont {A.~C.}\ \bibnamefont {{Bleszynski
  Jayich}}},\ }\href {\doibase 10.1038/ncomms5429} {\bibfield  {journal}
  {\bibinfo  {journal} {Nat. Commun.}\ }\textbf {\bibinfo {volume} {5}},\
  \bibinfo {pages} {4429} (\bibinfo {year} {2014})}\BibitemShut {NoStop}%
\end{thebibliography}

%

\end{document}